\documentclass[a4paper,11pt]{article}

\usepackage[hmargin=0.9in,vmargin=1in]{geometry}
\usepackage[T1]{fontenc}
\usepackage{lmodern}
\usepackage{microtype}
\usepackage{color}
\usepackage{dsfont}
\usepackage{extarrows}
\usepackage{enumerate}
\usepackage{amsmath}    
\usepackage{amssymb}
\usepackage{bbm}
\usepackage{graphicx}   
\usepackage{mathrsfs}
\usepackage{verbatim}   
\usepackage{subfigure}  

\newtheorem{theorem}{Theorem}[section]
\newtheorem{corollary}[theorem]{Corollary}
\newtheorem{lemma}[theorem]{Lemma}
\newtheorem{proposition}[theorem]{Proposition} 

\newtheorem{definition}[theorem]{Definition}
\newtheorem{remark}[theorem]{Remark}

\newcommand{\Tr}{{\rm Tr}}

\usepackage{cite}
\usepackage{algorithm,algorithmic}
\usepackage{textcomp}
\usepackage{hyperref}
\hypersetup{hidelinks}
\def\BibTeX{{\rm B\kern-.05em{\sc i\kern-.025em b}\kern-.08em
    T\kern-.1667em\lower.7ex\hbox{E}\kern-.125emX}}
\newcommand{\proof}{\par\medskip\noindent\textit{Proof: }\ignorespaces}

\newenvironment{IEEEkeywords}
  {\par\medskip\noindent\small\textbf{Keywords: }\ignorespaces}
  {\par\medskip}

\date{}
\begin{document}

\title{Stabilization of Time-Varying Perturbed Quantum Systems via Reduced Filters}
\author{Weichao Liang
\thanks{
{\small W. Liang is with the School of Automation Science and Engineering, Faculty of Electronic and Information Engineering, Xi’an Jiaotong University, Xi'an, 710049, Shaanxi, P.R. China (e-mail: weichao.liang@xjtu.edu.cn). }}
\and Daoyi Dong
\thanks{
{\small D. Dong is with Australian Artificial Intelligence Institute, Faculty of Engineering and Information Technology, University of Technology Sydney, Broadway, Ultimo, 2007, New South Wales, Australia, (daoyidong@gmail.com).}}
}

\maketitle

\begin{abstract}
In practical applications, quantum systems are inevitably subject to significant uncertainties, including unknown initial states, imprecise physical parameters, and unmodeled environmental noise, all of which pose major challenges to robust quantum feedback control. Under a quantum non-demolition measurement structure, we propose a feedback stabilization strategy based on a reduced quantum filter that achieves robustness against time-varying Hamiltonian perturbations and additional dissipative effects, without requiring prior knowledge of the initial state or exact system parameters. The controller propagates at most $N - 1$ independent real coordinates, compared with the $N^2 - 1$ real coordinates of a full density matrix, where $N$ is the dimension of the Hilbert space. This dimensional reduction is achieved without sacrificing the stabilization and robustness guarantees. Under the stated conditions, the analysis establishes almost-sure exponential convergence to the target subspace with an explicit sample exponent bound.
\end{abstract}




\section{Introduction}

Open quantum systems inevitably interact with their
environment, leading to dissipation and decoherence
that pose fundamental challenges to quantum control
and information processing
\cite{breuer2002theory,nielsen2010quantum}.
Continuous measurement provides real-time information
about the conditional quantum state, whose evolution
is described by stochastic master equations
\cite{belavkin1989nondemolition,bouten2007introduction,
barchielli2009quantum}. Measurement-based feedback
has been investigated for quantum state preparation,
coherence protection, quantum error correction,
and other applications
\cite{zhang2017quantum,vijay2012stabilizing,
ofek2016extending,mabuchi2009continuous}.  Recent research has also explored its role in connecting gravity and quantum matter~\cite{oppenheim2022constraints}.

Two challenges arise in the implementation of
SME-based feedback control. First, a full-state
filter for an $N$-dimensional quantum system
generally propagates $N^2-1$ independent real
parameters, making real-time feedback increasingly
demanding as the system dimension grows.
Various reduction methods have been developed,
including projection-based quantum filters
\cite{van2005quantum,nielsen2009quantum,
gao2019design,gao2020improved,amini2025feedback},
structure-preserving reduction of linear quantum
stochastic systems \cite{nurdin2014structures},
and exact observable-based reduction
\cite{grigoletto2025quantum}.
Second, unknown initial states, uncertain
physical parameters, and unmodeled environmental
interactions complicate feedback design and
stability analysis, particularly when Hamiltonian
and dissipative coefficients vary with time.
Robust feedback stabilization has been investigated
under different uncertainty structures
\cite{van2005feedback,mirrahimi2007stabilizing,
ticozzi2012stabilization,benoist2017exponential,
liang2025exploring,guatto2024improving}.

Our previous work
\cite{liang2024model,liang2025reduced}
introduced a reduced-filter feedback architecture
and investigated its stabilization properties
under model uncertainties and perturbations.
The present paper develops a more general
robustness analysis for bounded measurable
time-varying Hamiltonian, dissipative, and
measurement parameters. Under the 
quantum non-demolition measurement
structure, the perturbations
are required to preserve the target
subspace and may induce nontrivial dynamics
among the non-target sectors. This setting
goes beyond a straightforward time-dependent extension
of the preceding results: the coefficients are only assumed to be bounded and measurable in time, while pointwise accessibility and qualitative support properties alone do not yield the uniform recurrence estimates required for robust stabilization.

Our main contributions are summarized as follows:
\begin{enumerate}
\item \textbf{Structural conditions for target
accessibility:}
We establish a joint Hamiltonian--measurement
reachability condition \textbf{A6} and a 
noncancellation criterion \textbf{A8} that exclude
persistence in the set of states with zero target
population.
These conditions characterize the
operator-level accessibility mechanism. We also derive verifiable
sector-wise and Kalman-type sufficient
conditions (Proposition~\ref{prop:reach_certificates}) and construct a prediction
matrix compatible with the control Hamiltonian (Lemma~\ref{lem:hamiltonian_laplacian}).

\item \textbf{Recurrence under time-varying perturbations:}
We establish finite-mean entrance into
target neighborhoods uniformly over
the prescribed class of bounded measurable
coefficient paths. The proof combines
a logarithmic exit estimate, uniform
controlled preparation, and a support
argument exploiting the QND measurement
structure. Compactness methods and
time-translation-invariant estimates
extend the recurrence analysis to
nonautonomous dynamics
(Proposition~\ref{Prop:Recurrence}).

\item \textbf{Robust reduced-filter-based
exponential stabilization:}
We establish well-posedness of the coupled
system (Theorem~\ref{Prop:Invariance_q}) and prove robust feedback stabilization
without knowledge of the initial state or exact time-dependent parameters.
The controller propagates only $d$ independent
coordinates, instead of reconstructing
the full density matrix. Under the proposed
structural and measurement conditions,
the physical state and reduced filter
converge almost surely exponentially to
their joint target, with an explicit
sample exponent bound
(Theorem~\ref{Thm:GES Feedback}).
\end{enumerate}

The remainder of this paper is organized as follows: Section~\ref{Sec:SysDes} introduces the stochastic model and formulates the feedback stabilization problem. Section~\ref{sec:QSR} establishes exponential quantum state reduction for uncontrolled systems. Section~\ref{sec:ReducedFeedback} develops the reduced-filter-based feedback strategy and establishes its robustness under time-varying perturbations. Section~\ref{sec:simulation} presents a numerical example involving a three-level quantum system. Section~\ref{Sec:conclusion} concludes the paper and discusses future research directions.

\textbf{Notation.}
The imaginary unit is denoted by $\mathfrak{i}$. 
For $X\in\mathcal{B}(\mathcal{H})$, the Hilbert-space adjoint is $X^*$, where $\mathcal{B}(\mathcal{H})$ denotes the space of all linear operators on $\mathcal{H}$, and $\mathbf{I}$ denotes the identity operator.  Let
\(
\mathcal{B}_{*}(\mathcal{H}) := \{X\in\mathcal{B}(\mathcal{H}) \mid X=X^*\}.
\) 
The trace of $X\in\mathcal{B}(\mathcal{H})$ is $\mathrm{Tr}(X)$, and the Hilbert–Schmidt norm is  
\(
\|X\| := (\mathrm{Tr}(XX^*))^{1/2}.
\)  
For $v\in\mathcal{H}$, $\|v\|$ denotes the standard vector norm. 
The vector $\mathbf1$ has all entries equal to one, with dimension determined by context.  
For a finite-dimensional operator space $\mathcal E$, the space
$L^\infty([0,\infty);\mathcal E)$ consists of entrywise measurable
operator-valued functions satisfying
\(
\|X\|_{L^\infty}:=\textstyle\operatorname*{ess\,sup}_{t\geq0}\|X(t)\|<\infty.
\)
For $X,Y\in\mathcal{B}(\mathcal{H})$, $[X,Y]:=XY-YX$.  
For $x\in\mathbb{C}$, $\Re\{x\}$ denotes the real part. The space of continuously differentiable functions from $A$ to $B$ is denoted by $\mathcal C^1(A,B)$. On the closed simplex, $\mathcal C^1$ means the restriction of a continuously differentiable map on a relatively open neighborhood in its affine hull.
For a positive integer $m$, we write $[m]:=\{1,\dots,m\}$. If $\mathcal{H}=\mathcal{H}_S\oplus\mathcal{H}_R$ and $X\in\mathcal{B}(\mathcal{H})$, in a basis adapted to this decomposition $X$ has block form  
\(
X=\left[\begin{smallmatrix}
X_S & X_P\\
X_Q & X_R
\end{smallmatrix}\right],
\)  
where $X_S,X_R,X_P$ and $X_Q$ are matrices representing operators from $\mathcal{H}_S$ to $\mathcal{H}_S$, from $\mathcal{H}_R$ to $\mathcal{H}_R$, from $\mathcal{H}_R$ to $\mathcal{H}_S$, from $\mathcal{H}_S$ to $\mathcal{H}_R$, respectively.

\section{Stochastic dynamical model}
\label{Sec:SysDes}
We consider an open quantum system monitored through $m$ real diffusive measurement channels (homodyne channels or quadrature components of heterodyne detection) on an $N$-dimensional Hilbert space $\mathcal{H}$. 
The conditional state takes values in the set of density operators
\begin{align*}
\mathcal{S}(\mathcal{H}):=\{\rho\in\mathcal{B}(\mathcal{H})\mid\rho=\rho^*\geq 0,\mathrm{Tr}(\rho)=1\}.    
\end{align*}
Fix a filtered probability space
$(\Omega,\mathcal F,(\mathcal F_t)_{t\geq0},\mathbb P)$
supporting an $m$-dimensional
$(\mathcal F_t)$-Wiener process $W$.
The conditional state evolves according to the
following stochastic master equation (SME):
\begin{align}
d\rho(t)=&\mathcal{L}^u_{\gamma}(t,\rho(t))dt+\mathcal{P}(t,\rho(t))dt\nonumber\\
&\textstyle +\sum^m_{k=1}\sqrt{\theta_k(t)}\mathcal{G}_{L_k}(\rho(t))dW_k(t),\label{Eq:SME}\\
dY_k(t) =& \sqrt{\theta_k(t)}\Tr((L_k^*+L_k)\rho(t))dt + dW_k(t),\label{Eq:dY}
\end{align}
with $\rho(0)\in \mathcal{S}(\mathcal{H})$. Here, 
\begin{itemize}
    \item $\theta_k(t)=\eta_k(t)\gamma_k(t)\in[\underline{\theta}_k,\bar{\theta}_k]$ is a deterministic measurable function combining measurement efficiency $\eta_k(t)\in[\underline{\eta}_k,\bar{\eta}_k]\subset(0,1]$ and coupling strength $\gamma_k(t)\in[\underline{\gamma}_k,\bar{\gamma}_k]$ with $\underline{\gamma}_k>0$, associated with the $k$-th probe, where $\underline{\theta}_k=\underline{\eta}_k\underline{\gamma}_k$ and $\bar{\theta}_k=\bar{\eta}_k\bar{\gamma}_k$.
    \item The nominal Lindblad generator is 
    $$
    \mathcal{L}^{u}_{\gamma}(t,\rho):=-\mathfrak{i}[H_0(t)+u_t H_1,\rho]+\textstyle\sum^m_{k=1}\gamma_k(t)\mathcal{D}_{L_k}(\rho),
    $$ where $\mathcal{D}_{L_k}(\rho) := L_k \rho L_k ^* - \frac{1}{2}L_k^* L_k \rho - \frac{1}{2} \rho L_k^* L_k$.  Here $H_0(t),H_1\in\mathcal{B}_{*}(\mathcal{H})$ represent the free and control Hamiltonians, and the measurement operator $\sqrt{\gamma_k(t)}L_k\in\mathcal{B}(\mathcal{H})$ describes interaction with the $k$-th probe.
    \item The measurement back-action associated with $k$-th probe is described by 
    $$\mathcal{G}_{L_k}(\rho):=L_k\rho+\rho L^*_k-\Tr((L_k+L_k^*)\rho)\rho.$$
    \item Perturbations are modeled by 
    $$
    \mathcal{P}(t,\rho):=-\mathfrak{i}[\tilde{H}_0(t),\rho]+\textstyle\sum^{\bar{m}}_{r=1}\mathcal{D}_{C_r(t)}(\rho).
    $$
\end{itemize}
The time-dependent operators are deterministic functions $H_0,\tilde H_0\in L^\infty([0,\infty);\mathcal B_*(\mathcal H))$, and $C_r\in L^\infty([0,\infty);\mathcal B(\mathcal H))$ for all $r\in[\bar m]$,
while $H_1\in\mathcal B_*(\mathcal H)$ and $L_k\in\mathcal B(\mathcal H)$ are fixed. More explicitly, there exist finite bounds $M_0,M_{\mathrm p},M_{C,r}$ such that $\|H_0\|_{L^\infty}\leq M_0$, $\|\tilde H_0\|_{L^\infty}\leq M_{\mathrm p}$ and $\|C_r\|_{L^\infty}\leq M_{C,r}$.

The continuous semimartingale $Y_k(t)$ represents the measurement record of the $k$-th probe with quadratic variation $\langle Y_k(t)\rangle=t$. 
The control input $u_t$ is a bounded real-valued observation-adapted process. Existence, uniqueness, and invariance of solutions to~\eqref{Eq:SME} within $\mathcal{S}(\mathcal{H})$ follow arguments similar to those in~\cite[Section 3]{mirrahimi2007stabilizing} and~\cite[Chapter 5]{barchielli2009quantum}.

We now consider a decomposition: $\mathcal{H} = \mathcal{H}_0 \oplus \dots \oplus \mathcal{H}_d$, with orthogonal projections $\Pi_0, \dots, \Pi_d$ onto each subspace. Throughout, $d\geq1$ and every sector $\mathcal H_j$ is nonzero. 
We impose the following quantum non-demolition (QND) structure on the free Hamiltonian and the measurement operators:
\begin{align*}
    &\hspace{-2em} \textbf{A1:} &&
    H_0(t)=\mathrm{diag}[\mathsf{H}_0(t),\dots,\mathsf{H}_d(t)]\text{ with } \mathsf{H}_j\in L^\infty([0,\infty);\mathcal B_*(\mathcal H_j));\\
    &&& L_k=\textstyle\sum^{d}_{j=0}l_{k,j}\Pi_j \text{ with } l_{k,j}\in\mathbb{C}.
\end{align*}
Assumption~\textbf{A1} specifies a QND structure relative to the fixed sectors $\mathcal H_j$. The free Hamiltonian $H_0(t)$ preserves each sector, while every measurement operator $L_k$ acts as a scalar on it.
A physical example is continuous homodyne monitoring
of a spin system $J_z=\sum_j m_j\Pi_j$ with
$L=J_z$ and a free Hamiltonian
$H_0(t)=\omega(t)J_z$, as in
Section~\ref{sec:simulation}. Degenerate sectors
are permitted, and the free Hamiltonian may generate
nontrivial dynamics within each sector.
The feedback Hamiltonian and the perturbations
may  couple different sectors. Hence,
the complete controlled dynamics do not need to preserve the QND decomposition.

The QND structure is essential to the reduced-filter
construction and the stability analysis. In particular,
it yields the sector-wise measurement coefficients,
allows the QND projections to be used in the
kernel-propagation argument, and provides commuting
diffusion fields for the support analysis. 

For $j\in\{0,\dots,d\}$, write
$p_j(\rho):=\Tr(\Pi_j\rho)$ and define 
\begin{align*}
&\mathbf{d}_j(\rho):=\|\rho-\Pi_j\rho \Pi_j\|,\\
&\mathbf{B}_{r}(\mathcal{H}_j):=\{\rho\in\mathcal{S}(\mathcal{H})\mid \mathbf{d}_j(\rho)<r\},\\
&\mathcal{I}(\mathcal{H}_j):=\{\rho\in\mathcal{S}(\mathcal{H})\mid\mathrm{Tr}(\Pi_j\rho)=1\}.
\end{align*}
\begin{definition}
The subspace $\mathcal{H}_j$ is called almost surely invariant  if, for all $\rho(0)\in \mathcal{I}(\mathcal{H}_j)$, $\rho(t)\in \mathcal{I}(\mathcal{H}_j)$ for all $t>0$ almost surely.
\end{definition}


\textbf{Problem setting:} We study feedback stabilization of~\eqref{Eq:SME} toward the target subspace $\mathcal{H}_0$ using a reduced filter instead of full state estimation. For consistency with the block notation, set
$\mathcal H_S=\mathcal H_0$, $\mathcal H_R=\mathcal H_0^\perp$, and let
$\Pi_0^\perp=\mathbf I-\Pi_0$. The blocks $\rho_R$ and $C_{r,P}$ are
understood relative to $\mathcal H_0\oplus\mathcal H_R$.

Throughout, we focus on perturbations that preserve the invariance of the target subspace $\mathcal{H}_0$. Intuitively, it means that the perturbations do not induce transitions that drive the state of system out of $\mathcal{H}_0$, and then stabilization towards this subspace remains feasible. Following~\cite{ticozzi2008quantum}, we formalize this assumption as:
\begin{description}
\item[\textbf{A2:}] For almost every $t\geq0$,
$\tilde H_{0,P}(t)=0$, and for $r\in[\bar m]$,
$C_{r,Q}(t)=0$ with
$\sum_{r}C_{r,S}(t)^*C_{r,P}(t)=0$.
\end{description}
Here, $\tilde{H}_{0,P}(t)$ denotes the coupling between the subspace $\mathcal{H}_0$ and its orthogonal complement $\bigoplus^d_{i=1}\mathcal{H}_i$ in the perturbed Hamiltonian, while $C_{r,Q}(t)$ and $C_{r,P}(t)$ represent, respectively, the off-diagonal blocks of the noise operators that connect $\mathcal{H}_0$ with the external subspaces.
\textbf{A2} provides a sufficient condition for target invariance under the uncontrolled dynamics. It permits Hamiltonian evolution and dissipative transfer into $\mathcal H_0$, while preventing perturbation-induced leakage from it. The related generator is given by
\[
\operatorname{Tr}(\Pi_0\mathcal P(t,\rho)) = \textstyle \sum_{r=1}^{\bar m} \operatorname{Tr} \big(C_{r,P}(t)\rho_R C_{r,P}(t)^*\big) \geq 0,
\]
where $\rho_R=\Pi_R\rho\Pi_R|_{\mathcal H_R}$.

\section{Stochastic evolution in the QND decomposition}
\label{sec:QSR}
We first analyze the uncontrolled dynamics ($u_t\equiv 0$) to identify the state components that govern large-time behavior. In the QND setting, the diagonal weights $p_i(t):=\Tr(\rho(t)\Pi_i)$ carry the relevant asymptotic information; see also~\cite{benoist2014large}.

\textbf{A2} ensures only the invariance of the target sector. For the QND reduction result below, we impose the explicit sector-preserving condition:
\begin{align*}
    &\hspace{-2em} \textbf{A-qsr:} &&
    \tilde{H}_0(t)=\mathrm{diag}[\tilde{\mathsf{H}}_0(t),\dots,\tilde{\mathsf{H}}_{d}(t)]\text{ with } \tilde{\mathsf{H}}_j\in L^\infty([0,\infty);\mathcal B_*(\mathcal H_j));\\
    &&& C_r(t)=\mathrm{diag}[C_{r,0}(t),\dots,C_{r,d}(t)]\text{ with }C_{r,j}\in L^\infty([0,\infty);\mathcal B(\mathcal H_j)).
\end{align*}
Assumption  \textbf{A-qsr} implies \textbf{A2}. Moreover, we impose the following identifiability assumption:
\begin{description}
    \item[\textbf{A3:}]For all $i\neq j$, there exists at least one $k\in[m]$ such that $\Re\{l_{k,i}\}\neq \Re\{l_{k,j}\}$.
\end{description}
Define the uniform measurement-separation rate
\begin{equation*}
\textstyle \mathfrak E_l:=\min_{i\neq j}\sum_{k=1}^{m}\underline{\theta}_k(\Re\{l_{k,i}\}-\Re\{l_{k,j}\})^2.
\end{equation*}
Since $\underline{\theta}_k>0$ for all $k\in[m]$, \textbf{A3} implies $\mathfrak E_l>0$.
To quantify dispersion among the QND sector populations, define
\begin{equation*}
V_{\rm QSR}(\rho) := \sum_{0\leq i<j\leq d} \sqrt{p_i(\rho)p_j(\rho)}.
\end{equation*}
Then, $V_{\rm QSR}\geq0$ and $V_{\rm QSR}(\rho)=0$ if and only if $\rho\in\bigcup_{j=0}^{d}\mathcal I(\mathcal H_j).$ By applying It\^o's formula on the invariant relative interior determined by the initially nonzero sector populations~\cite[Theorem 5.1]{liang2019exponential}, we have 
$$\mathscr L_t V_{\rm QSR}(\rho) \leq -\frac{\mathfrak E_l}{2} V_{\rm QSR}(\rho).$$
Hence, the standard QND state-reduction arguments in \cite{liang2019exponential,cardona2020exponential,liang2024model} extend directly to the bounded time-varying setting.
\begin{theorem}
\label{Thm:QSR}
Suppose that $u_t\equiv0$ and that \textbf{A1}, \textbf{A3}, and \textbf{A-qsr} hold. Then, for every $\rho(0)\in\mathcal S(\mathcal H)$,
\begin{align*}
    &\mathbb E[V_{\rm QSR}(\rho(t))]\leq e^{-\mathfrak E_l t/2}V_{\rm QSR}(\rho(0)), \quad t\geq0,\\
    &\limsup_{t\to\infty}\frac{1}{t}\log V_{\rm QSR}(\rho(t))\leq -\frac{\mathfrak E_l}{2}, \quad a.s.
\end{align*}
Moreover, there exists a random variable $\mathfrak J_\infty\in\{0,\ldots,d\}$ such that
\begin{align*}
    &p_j(t) \to \mathds{1}_{\{\mathfrak J_\infty=j\}}, \quad \forall j=0,\ldots,d, \quad a.s., \\
    &\mathbb P(\mathfrak J_\infty=j)= p_j(0) = \Tr(\Pi_j\rho(0)), \quad \forall j=0,\ldots,d.
\end{align*}
\end{theorem}

\medskip

Theorem~\ref{Thm:QSR} shows that the asymptotic QND reduction is characterized by the population vector $(p_0,\dots,p_d)$.  On the event that $\mathcal{I}(\mathcal H_j)$ is selected, $p_j\to1$ and $p_i\to0$ for $i\neq j$. Due to $\rho(t)\geq 0$,  $\Pi_i\rho(t)\Pi_j$ for $i\neq j$ vanish asymptotically. This motivates the sector populations as reduced information variables: while a $N$-dimensional density matrix has $N^2-1$ independent real parameters, the population vector has only $d$ independent coordinates since $\sum_{j=0}^{d}p_j=1$.

Theorem~\ref{Thm:QSR} also illustrates the role of the QND
decomposition in selecting stabilization targets.
Under \textbf{A-qsr}, the uncontrolled dynamics
randomly selects one of the invariant sectors,
whereas feedback control aims to stabilize a
prescribed target sector $\mathcal H_0$.
The latter requires additional reachability and
robustness conditions, particularly when perturbations
do not preserve the individual non-target sectors.
However, this reduction does not imply that the sector populations form a closed system with feedback.  Indeed, the feedback contribution to their dynamics contains $-\mathfrak i u_t \Tr(\Pi_j[H_1,\rho(t)])$ which generally depends on the coherences of $\rho(t)$. Accordingly, the reduced filter introduced in Section~\ref{sec:ReducedFeedback} is not simply the diagonal part of a full quantum filter. It provides a low-dimensional stochastic controller state designed to keep the sector-level measurement information relevant to target selection without reconstructing the full density matrix.

%

\section{Stabilization by the reduced filter}
\label{sec:ReducedFeedback}

In this section, we explore the robustness of the state feedback stabilization strategy proposed in~\cite{liang2024model} for system~\eqref{Eq:SME}, subject to perturbations $\mathcal P(t,\rho)$. Throughout, we  assume that the measurement operators $L_k$ are known, while the parameters $\gamma_k(t)$ and $\eta_k(t)$ may not be precisely identified. This assumption can be interpreted as allowing proportional time-varying perturbations in $L_k$.  Importantly, this  assumption plays a crucial role in deriving the main result, as it emphasizes that accurate knowledge of the measurement operators is essential for the robust stability.

Due to relation~\eqref{Eq:dY}, we rewrite system~\eqref{Eq:SME} as follows:
\begin{align}
d\rho(t)=&\mathcal{L}^u_{\gamma}(t,\rho(t))dt+\mathcal{P}(t,\rho(t))dt+\sum_{k}\sqrt{\theta_k(t)}\mathcal{G}_{L_k}(\rho(t))\nonumber\\
&\times\big(dY_k(t)-\sqrt{\theta_k(t)}\Tr((L_k^*+L_k)\rho(t))dt\big),\label{Eq:SME_Y}
\end{align}
with $\rho(0)\in\mathcal{S}(\mathcal{H})$.
From a practical perspective, the initial state, the free Hamiltonian, the model parameters and the perturbation cannot be precisely known. For comparison, a full-state  filter driven by the same record has the form~\cite{liang2025exploring}:
\begin{align}
d\hat{\rho}(t)=&\widehat{\mathcal L}^{\,u}(\hat\rho(t))dt+\sum^m_{k=1}\sqrt{\hat{\theta}_k}\mathcal{G}_{L_k}(\hat{\rho}(t))\left(dY_k(t)-\sqrt{\hat{\theta}_k}\Tr((L_k^*+L_k)\hat{\rho}(t))dt\right),\quad \hat\rho(0)\in\mathcal S(\mathcal H),\label{Eq:Filter_Y}
\end{align}
where
\(
\widehat{\mathcal L}^{\,u}(\hat\rho)
:=-\mathfrak i[\hat H_0+uH_1,\hat\rho]
       +\sum_{k=1}^m\hat\gamma_k\mathcal D_{L_k}(\hat\rho),
\)
$\hat H_0=\operatorname{diag}[\hat{\mathsf H}_0,\ldots,\hat{\mathsf H}_d]$
with fixed $\hat{\mathsf H}_j\in\mathcal B_*(\mathcal H_j)$,
 and $\hat{\theta}_k=\hat{\eta}_k\hat{\gamma}_k$ with $\hat{\gamma}_k\in[\underline{\gamma}_k,\bar{\gamma}_k]$, $\hat{\eta}_k\in[\underline{\eta}_k,\bar{\eta}_k]$. Unlike the true system, which may involve time-varying $H_0(t)$ and $\theta_k(t)$, the estimator uses fixed values selected from prior information, reducing online computational complexity. A full-state feedback law would require real-time integration of~\eqref{Eq:Filter_Y}.

\subsection{Motivation for a reduced filter}

As the system dimension increases, integrating the full filter~\eqref{Eq:Filter_Y} in real time becomes computationally demanding. This motivates the design of controllers that exploit only partial information rather than full state estimation.

Before introducing the reduced filter, we distinguish a controller
state from an exact conditional population filter. Even under
\textbf{A1}, the controlled population drift contains
$-\mathfrak i u_t\Tr([H_1,\rho]\Pi_n)$ and may depend on coherences.
Perturbations allowed by \textbf{A2} may also transfer mass between
non-target sectors. Thus, a full filter $\hat \rho$ remains
$N^2-1$ dimensional. The reduced filter instead replaces the unclosed drift with a Markov-generator term while exploiting the QND measurement structure. It is designed for stabilization rather than consistent estimation of all sector populations.


Since  $L_k$ are diagonal in the QND basis, their contribution to the measurement back-action can be simplified to:
\begin{align}
    \Tr(\mathcal{G}_{L_k}(\rho)\Pi_n)&=2p_n\big(\Re\{l_{k,n}\}-\textstyle\sum^{d}_{j=0}\Re\{l_{k,j}\}p_j\big).\label{Eq:Diag}
\end{align}
Motivated by this structure, and by the analysis in Section~\ref{sec:QSR}, we introduce a reduced-order estimator whose state $\hat{q}(t)\in\mathbb{R}^{d+1}$ estimates sector weights through the measurement record. Inspired by~\cite{cardona2020exponential,liang2024model}, we propose the following observation-driven stochastic differential equation (SDE):
\begin{align}
d \hat{q}_n(t)=& u_t\sum^d_{j=0}\Gamma_{n,j}\hat{q}_j(t)dt+2\hat{q}_{n}(t)\sum^m_{k=1}\sqrt{\hat{\theta}_k}\Phi^k_n(\hat{q}(t))\left(dY_k(t)-2\sqrt{\hat{\theta}_k}\Lambda_k(\hat{q}(t))dt\right),  \label{Eq:SDE_Y} 
\end{align}
where $\Phi^k_n(\hat{q}):=\Re\{l_{k,n}\}-\Lambda_k(\hat{q})$ with $\Lambda_k(\hat{q}):=\textstyle\sum^{d}_{j=0}\Re\{l_{k,j}\}\hat{q}_j,$
 $\hat{q}_n=e_n^*\hat{q}$, and $\{e_0,\dots,e_d\}$ is the canonical coordinate basis of $\mathbb{R}^{d+1}$. $u_t:=u(\hat{q}(t))$ represents the feedback controller.

Moreover, we assume that the matrix $\Gamma \in \mathbb{R}^{(d+1)\times (d+1)}$ satisfies the following condition:
\begin{description}
\item[\textbf{A4:}]
\begin{enumerate}[(i)]
\item $\sum_{i=0}^d\Gamma_{i,j}=0$, $\Gamma_{j,j}<0$ for every $j$,
and $\Gamma_{i,j}\geq0$ for $i\ne j$.
\item For every nonempty $I\subseteq[d]$, there are $i\in I$ and
$j\in\{0,\ldots,d\}\setminus I$ such that $\Gamma_{j,i}>0$.
\end{enumerate}
\end{description}
Part (i) makes  $\Gamma$ a Markov generator in the column convention, part (ii)
ensures that every non-target vertex has a directed path to $0$.

The observation-driven term in~\eqref{Eq:SDE_Y} exploits the sector-wise measurement structure in~\eqref{Eq:Diag}, while the drift $u_t\sum_{j=0}^d\Gamma_{n,j}\hat q_j$ is introduced to model population transfer under \textbf{A4}(i), rather than to reproduce the exact Hamiltonian drift.

\subsection{Stochastic model for the reduced filter}
Based on~\eqref{Eq:dY}, we rewrite~\eqref{Eq:SDE_Y} as the following Wiener process-driven SDE for $n\in\{0,\dots,d\}$, 
\begin{align}
d\hat{q}_n(t)=& u_t\sum^d_{j=0}\Gamma_{n,j}\hat{q}_j(t)dt+2\hat{q}_{n}(t)\sum^m_{k=1}\sqrt{\hat{\theta}_k}\Phi^k_n(\hat{q}(t))\left(dW_k(t)+\mathfrak{T}_k(t,\rho(t),\hat{q}(t))dt\right),\label{Eq:SDE}
\end{align}
where 
$\textstyle \mathfrak{T}_k(t,\rho,\hat{q}):=\sqrt{\theta_k(t)}\mathrm{Tr}((L_k+L^*_k)\rho)-2\sqrt{\hat{\theta}_k}\Lambda_k(\hat{q}),$ and $\rho(t)$ is the solution to SME~\eqref{Eq:SME}.
Define the probability simplex
\[
\bar{\mathcal O}_d
:=\{
\hat q\in[0,1]^{d+1}
\,|\,
\textstyle\sum_{n=0}^d\hat q_n=1
\}
\]
and its relative interior
$\mathcal O_d
:=\{\hat q\in\bar{\mathcal O}_d
\mid \hat q_n>0,\ n=0,\ldots,d\}$.
The feedback controller $u_t:=u(\hat{q}(t))$ satisfies the following assumption:
\begin{description}
\item[\textbf{A5:}] 
$u\in\mathcal C^1(\bar{\mathcal O}_{d},[0,\infty))$,
$u(e_0)=0$, and $u(\hat q)>0$ for $\hat q\ne e_0$.
There exist $\alpha>0$ and $c_u,r_u>0$ such that $u(\hat q)\leq c_u(1-\hat q_0)^{1+\alpha}$ if $1-\hat q_0<r_u$.
\end{description}
The $\mathcal C^1$ regularity gives the state-space Lipschitz
bounds used for well-posedness and the support lemma. The separate
power bound ensures $u(\hat q)/(1-\hat q_0)\to0$, as used in the local target-stability estimate.  For example, the feedback law
\begin{equation}\label{Eq:u}
u(\hat q)=a(1-\hat q_0)^b,
\quad a>0,\quad b>1,
\end{equation}
satisfies \textbf{A5} with
$0<\alpha\leq b-1$.

Let $\mathcal X:=\mathcal S(\mathcal H)\times\bar{\mathcal O}_{d}$ and write $Z(t)=(\rho(t),\hat q(t))$.
We first establish well-posedness of the coupled system and invariance of the state space.
\begin{theorem}\label{Prop:Invariance_q}
Suppose \textbf{A4}(i) and \textbf{A5}  hold. For every initial condition in
$\mathcal X$,
the coupled system~\eqref{Eq:SME} and \eqref{Eq:SDE} has a unique global strong
solution  $Z(t)\in\mathcal X$. If $\hat q(0)\in\mathcal O_{d}$, then
$\hat q(t)\in\mathcal O_{d}$ for all  $t\geq0$ almost surely.
\end{theorem}

The proof is given in Appendix~\ref{App:ProofThmInv_q}.
For each fixed admissible deterministic coefficient path,
the uniform Lipschitz bounds yield a Feller
evolution family, while pathwise uniqueness and the
independence of Wiener increments ensure that the
time-augmented process $(t,Z(t))$ is strong Markov.

\subsection{Feedback stabilization to target subspaces}
We adapt methods from~\cite{liang2024model,liang2025exploring} to establish stability of the joint target and almost-sure exponential convergence using the reduced filter. 
%

\subsubsection{Joint Hamiltonian--measurement reachability}
We first formulate a structural condition involving only the control Hamiltonian $H_1$ and the QND sector projections $\{\Pi_n\}^d_{n=0}$. Define a sequence of subspaces
\begin{equation}\label{eq:reach_sequence}
\mathcal R_0=\mathcal H_0,\quad \textstyle 
\mathcal R_{\ell+1}=\mathcal R_\ell+
       \big(\sum_{n=0}^d\Pi_nH_1\mathcal R_\ell\big).
\end{equation}
Let $J$ be the first index with $\mathcal R_{J+1}=\mathcal R_J$. Since $\mathcal{H}$ is finite dimensional, $J\leq N-\dim\mathcal H_0$. Set $\mathcal R_\infty=\mathcal R_J$. Define $$\mathcal I_0:=\{\rho\in\mathcal S(\mathcal H)|\Tr(\rho \Pi_0)=0\}.$$ The following reachability condition, together
with \textbf{A8}, will be used to establish non-persistence in $\mathcal{I}_0$. 
\begin{description}
\item[\textbf{A6:}] $\mathcal R_\infty=\mathcal H$.
\end{description}

\begin{proposition}
\label{prop:joint_invariance}
The space $\mathcal R_\infty$ is the smallest subspace containing
$\mathcal H_0$ that is invariant under $H_1$ and all $\Pi_n$.
Then, \textbf{A6} is equivalent to the absence of a
nonzero subspace $\mathcal V\subseteq\mathcal H_0^\perp$ satisfying $H_1\mathcal V\subseteq\mathcal V$ and $\Pi_n\mathcal V\subseteq\mathcal V$ for all $n\in\{0,\dots, d\}$.
\end{proposition}
\proof
We first show that $\mathcal R_\infty$ is invariant under
$H_1$ and all the projections $\Pi_n$.
Since $\mathcal R_0=\mathcal H_0$, then $\Pi_n\mathcal R_0\subseteq\mathcal R_0$ for every
$n\in\{0,\ldots,d\}$.
Suppose that $\mathcal R_\ell$ is invariant under
all $\Pi_n$. Then, by~\eqref{eq:reach_sequence}, for every $i\in\{0,\ldots,d\}$, we have
\begin{align*}
\Pi_i\mathcal R_{\ell+1}
&= \Pi_i\mathcal R_\ell +\sum_{n=0}^d\Pi_i\Pi_n H_1\mathcal R_\ell=\Pi_i\mathcal R_\ell +\Pi_i H_1\mathcal R_\ell\subseteq\mathcal R_{\ell+1},
\end{align*}
Thus, each $\mathcal R_\ell$ is invariant under every $\Pi_n$.

Since  $\mathcal R_\ell \subseteq\mathcal R_{\ell+1}$ and $\mathcal H$ is finite dimensional,
there exists $J\in\mathbb{N}$ such that
$\mathcal R_{J+1}=\mathcal R_J=\mathcal R_\infty$.
It follows from~\eqref{eq:reach_sequence} that, for all $n$, $\Pi_nH_1\mathcal R_\infty
\subseteq\mathcal R_\infty$. Thus, we deduce
\begin{equation}\label{eq:invariant R_infty}
\textstyle H_1\mathcal R_\infty \subseteq \sum_{n=0}^d\Pi_nH_1\mathcal R_\infty
\subseteq\mathcal R_\infty.
\end{equation}
It implies that $\mathcal R_\infty$ is invariant under $H_1$ and all $\Pi_n$.

Let $\mathcal W$ be any subspace containing $\mathcal H_0$ and invariant under $H_1$ and all $\Pi_n$. Next, we show that $\mathcal R_\ell\subseteq\mathcal W$ for every $\ell\geq0$. The inclusion holds trivially for $\ell=0$.  If $\mathcal R_\ell\subseteq\mathcal W$, 
\[
\Pi_nH_1\mathcal R_\ell
\subseteq\Pi_nH_1\mathcal W
\subseteq\mathcal W, \quad \forall n\in\{0,\dots,d\}.
\] 
Then, by~\eqref{eq:reach_sequence}, we have  $\mathcal R_{\ell+1}\subseteq\mathcal W.$ It follows by induction that $\mathcal R_\infty\subseteq\mathcal W$. Therefore, $\mathcal R_\infty$ is the smallest common invariant subspace containing $\mathcal H_0$.

Finally, we establish the equivalence. Suppose that
$\mathcal R_\infty\neq\mathcal H$, and set
$\mathcal V=\mathcal R_\infty^\perp$.
Then, $\mathcal V\subseteq\mathcal H_0^\perp$ since $\mathcal H_0\subseteq\mathcal R_\infty$. Due to the relation~\eqref{eq:invariant R_infty}, for any $v\in\mathcal V$ and
$w\in\mathcal R_\infty$, $\langle H_1v|w\rangle =\langle v|H_1w\rangle=0$.
Hence, $H_1v\in\mathcal V$.
Similarly, since $\Pi_n\mathcal R_\infty\subseteq \mathcal R_\infty$, we have
$\Pi_nv\in\mathcal V$ for all $n$.
Thus, $\mathcal V$ satisfies the stated
invariance conditions.

Conversely, suppose that there exists a nonzero
subspace $\mathcal V\subseteq\mathcal H_0^\perp$
that is invariant under $H_1$ and all $\Pi_n$.
Since $H_1$ and the $\Pi_n$ are self-adjoint, $\mathcal V^\perp$ is also invariant under them.  Moreover,
$\mathcal H_0\subseteq\mathcal V^\perp$.
By the results established above, we deduce $\mathcal R_\infty
\subseteq\mathcal V^\perp
\neq\mathcal H,$
where the inequality follows from $\mathcal V\neq\{0\}$.
Therefore, $\mathcal R_\infty\neq\mathcal H$,
which completes the proof.
\hfill$\square$

\smallskip

Assumption~\textbf{A6} excludes any nonzero subspace of $\mathcal H_0^\perp$ that is invariant under both $H_1$ and $\{\Pi_n\}^d_{n=0}$. Under \textbf{A1} and \textbf{A3}, these projections are determined by the joint spectral structure of the measurement operators $L_k$. Hence, the condition incorporates both the
control Hamiltonian and the measurement structure, allowing Hamiltonian-dark directions
that are not preserved by the QND projections.

The following sector and Kalman tests provide convenient sufficient conditions for \textbf{A6}. Let $\mathscr E_\lambda(T)$ denote the eigenspace of $T$ and write $H_{1,nn}=\Pi_nH_1\Pi_n|_{\mathcal H_n}$.
The \emph{sector test} is
\begin{equation}\label{eq:sector_test}
\textstyle \bigcap_{i\ne j}\ker(\Pi_iH_1\Pi_j)
 \cap\mathscr E_\lambda(H_{1,jj})=\{0\},
\end{equation}
for every $j\in[d]$ and every eigenvalue $\lambda$ of the indicated
compression, and  all kernels are taken in $\mathcal H_j$.
This is a sector-wise Hautus test \cite[Sec.~3.3]{songtag1998control}.

\begin{lemma}\label{Lem:non_invariance_I_n}
Condition~\eqref{eq:sector_test} holds if and only if
$[H_1,\rho]\ne0$ for every $\rho\in\mathcal I(\mathcal H_n)$ and
every $n\in[d]$.
\end{lemma}
\proof
Fix $n>0$ and $\rho\in\mathcal I(\mathcal H_n)$. Since $\rho=\Pi_n\rho\Pi_n$, $\Pi_n[H_1,\rho]\Pi_n=[H_{1,nn},\rho]$, $\Pi_i[H_1,\rho]\Pi_n=\Pi_iH_1\Pi_n\rho$ and $\Pi_n[H_1,\rho]\Pi_i=-\rho\Pi_nH_1\Pi_i$ for $i\neq n$,
and all blocks with both indices different from $n$ vanish.
Thus, $[H_1,\rho]=0$ if and only if $[H_{1,nn},\rho]=0$ and $\Pi_iH_1\Pi_n\rho=0$ for $i\neq n$.
If these equalities hold,  $\operatorname{Ran}(\rho)$
is invariant under $H_{1,nn}$ and is contained in every
$\ker(\Pi_iH_1\Pi_n)$ for  $i\ne n$. 
This is a nonzero invariant subspace because $\Tr(\rho)=1$.  Since $H_{1,nn}|_{\operatorname{Ran}(\rho)}$ is Hermitian, it has a unit eigenvector $w$ with some eigenvalue $\lambda$.
This $w$ violates~\eqref{eq:sector_test}.
Conversely, if a nonzero $w$ belongs to the intersection in
\eqref{eq:sector_test}, then $H_1w=\lambda w$.
Its normalized projector $\rho=|w\rangle\langle w|/\|w\|^2$
belongs to $\mathcal I(\mathcal H_n)$ and commutes with $H_1$.
Both implications hold for each $n>0$.\hfill$\square$

An immediate consequence is the following easily verifiable sufficient condition.
\begin{corollary}\label{cor:H1}
The sector test~\eqref{eq:sector_test} holds if
$(\mathbf I-\Pi_n)H_1\Pi_n:\mathcal H_n\to\mathcal H_n^\perp$
is injective for every $n\in[d]$.
\end{corollary}

The \emph{Kalman rank condition} is: for some integer $\nu\geq0$,
\begin{equation}\label{eq:rank_test}
\operatorname{rank}[H_{1,Q},H_{1,R} H_{1,Q},\ldots,H_{1,R}^\nu H_{1,Q}]
       \geq\dim \mathcal H_0^\perp-1.
\end{equation}
The full-rank condition is
\begin{equation}\label{eq:full_rank_test}
\operatorname{rank}[H_{1,Q},H_{1,R} H_{1,Q},\ldots,H_{1,R}^\nu H_{1,Q}]=\dim \mathcal H_0^\perp.
\end{equation}

\begin{proposition}
\label{prop:reach_certificates}
Assumption~\textbf{A6} holds if either the full-rank
condition~\eqref{eq:full_rank_test} is satisfied, or
the rank condition~\eqref{eq:rank_test} holds together
with the sector test~\eqref{eq:sector_test}.
Moreover, if all QND sectors are one-dimensional,
\textbf{A6} holds if and only if the undirected graph
with vertex set $\{0,\ldots,d\}$ and edge set
\(
\{\{i,j\}:0\leq i<j\leq d,
\Pi_iH_1\Pi_j\neq0\}
\)
is connected.
\end{proposition}

\proof
By Proposition~\ref{prop:joint_invariance}, $\mathcal R_\infty$ contains
$\mathcal H_0$ and is invariant under $H_1$ and $\mathbf I-\Pi_0$.
For $w\in\mathcal H_0$, $H_{1,Q} w=(\mathbf I-\Pi_0)H_1w\in\mathcal R_{\infty}$, which implies  $H_{1,Q} w\in\mathcal R_\infty\cap\mathcal H_0^\perp$.
If $H_{1,R} ^r H_{1,Q} w\in\mathcal R_\infty\cap\mathcal H_0^\perp$, then
$$H_{1,R}^{r+1} H_{1,Q} w=(\mathbf I-\Pi_0) H_1 H_{1,R}^r H_{1,Q} w\in\mathcal R_\infty\cap\mathcal H_0^\perp.$$
By the induction, we obtain
\[
\mathcal H_0\oplus\operatorname{span}_{0\leq r\leq\nu}
       \operatorname{Ran}(H_{1,R}^r H_{1,Q})\subseteq\mathcal R_\infty,
\]
where $\operatorname{Ran}(H_{1,R}^r H_{1,Q})\subseteq\mathcal H_0^\perp$. Under~\eqref{eq:rank_test}, it follows that
$\dim\mathcal R_\infty\geq\dim\mathcal H_0+(\dim\mathcal H_0^\perp-1)=N-1$.
If $\mathcal R_\infty\ne\mathcal H$, then $\mathcal R_\infty^{\perp}=\operatorname{span}\{w\}\subseteq\mathcal H_0^\perp$  for some $w\ne0$.
All $\Pi_n$ and $H_1$ preserve this line due to their self-adjointness and Proposition~\ref{prop:joint_invariance}.

Write $\Pi_nw=c_nw$ where $c_n\in\{0,1\}$ due to $\Pi_n^2=\Pi_n$. Since $\sum_i \Pi_i=\mathbf I$, we have $\sum_i c_i=1$. Thus, there exists exactly one $c_n=1$ with $n\geq 1$ since $w\perp\mathcal H_0$. Hence, $w\in\mathcal H_n$ and $H_1w=\lambda w$ for some real $\lambda$, contradicting \eqref{eq:sector_test}. Thus, $\mathcal R_\infty=\mathcal H$. Under~\eqref{eq:full_rank_test}, the displayed inclusion already has
dimension $N$, which implies the same conclusion directly.

For one-dimensional sectors, each $\mathcal R_\ell$ is the span of
the vertices obtained after at most $\ell$ graph steps from zero:
this follows by induction from~\eqref{eq:reach_sequence}, since a
nonzero projected image spans its whole one-dimensional sector.
The recursion reaches every sector exactly when the graph is connected.
\hfill$\square$

The subtraction of one in~\eqref{eq:rank_test} allows a possible
one-dimensional invariant complement, while the sector test removes that
last possibility.

The coupling structure of $H_1$ also provides
an explicit construction of the prediction matrix
$\Gamma$ satisfying the required connectivity conditions \textbf{A4}.

\begin{lemma}
\label{lem:hamiltonian_laplacian}
Assume \textbf{A6}. For $\kappa>0$, define
\[
\Gamma_{ij}=\kappa\|\Pi_iH_1\Pi_j\|^2,~ \forall i\ne j,\quad \textstyle \Gamma_{jj}=-\sum_{i\ne j}\Gamma_{ij}.
\]
Then, $\Gamma$ is a symmetric irreducible Markov generator satisfying \textbf{A4} with
$\ker(\Gamma)=\operatorname{span}\{\mathbf 1\}$.
\end{lemma}
\proof
Since $H_1$ is Hermitian, $\Gamma$ is symmetric with nonnegative off-diagonal entries and zero column sums.
Suppose that the associated graph is disconnected, and let $I\subseteq[d]$ be a connected component. Since $\Pi_iH_1\Pi_j=0$ whenever $i\in I$ and $j\notin I$, the nonzero subspace
\(
\bigoplus_{j\in I}\mathcal H_j
\subseteq\mathcal H_R
\)
is invariant under $H_1$ and all $\Pi_n$, which contradicts \textbf{A6} by Proposition~\ref{prop:joint_invariance}.
Thus, the graph is connected, and $\Gamma$
is irreducible and satisfies \textbf{A4}(ii).
Moreover, $\Gamma_{jj}<0$ for every $j$,
and \textbf{A4}(i) follows by construction.
Finally, for every $x\in\mathbb R^{d+1}$,
\[
-x^\top\Gamma x =\textstyle \frac{\kappa}{2}\sum_{i\ne j} \|\Pi_iH_1\Pi_j\|^2(x_i-x_j)^2,
\]
which vanishes if and only if $x$ is constant. Hence,
$\ker(\Gamma)=\operatorname{span}\{\mathbf 1\}$.
\hfill$\square$

\subsubsection{Exit from incorrect target estimates}

Under \textbf{A2}, perturbations may transfer population among
non-target sectors. Thus, we estimate exit from configurations with small target population and an estimate near $e_0$. 
To quantify the effect of time-varying measurement perturbations, we introduce the normalized parameter 
$$
\chi_k(t):= (\theta_k(t)/\hat{\theta}_k)^{1/2}\in [\underline{\chi}_k,\bar{\chi}_k],
$$
where $\underline{\chi}_k=(\underline{\theta}_k/\hat{\theta}_k)^{1/2}$ and $\bar{\chi}_k=(\bar{\theta}_k/\hat{\theta}_k)^{1/2}$ for all $k\in [m]$.
For $\chi=(\chi_1,\ldots,\chi_m)$ and $n,j\in[d]$, define
\begin{align}
G_{nj}(\chi)
:=&\textstyle 2\sum_{k=1}^{m}\hat\theta_k
\bigl(\Re\{l_{k,n}\}-\Re\{l_{k,0}\}\bigr)
\bigl(2\chi_k\Re\{l_{k,j}\}
-\Re\{l_{k,n}\}-\Re\{l_{k,0}\}\bigr).
\label{eq:G_nj}
\end{align}
Then, we impose the following assumption:
\begin{description}
\item[\textbf{A7:}] There exist $\beta_n\geq0$,
$\sum_{n=1}^d\beta_n=1$, and $\kappa_R>0$ such that, for every $j\in[d]$ and every admissible parameter vector $\chi$, $\sum_{n=1}^d\beta_nG_{nj}(\chi) \geq\kappa_R.$
\end{description}
Assumption \textbf{A7} ensures a uniform negative bound
on the measurement contribution to the generator of
$W_\beta$ near configurations with vanishing target
population and $\hat q$ close to $e_0$.
Since $G_{nj}(\chi)$ is affine in $\chi$, this condition
can be verified at the vertices of the admissible
parameter box, reducing its verification to a finite
linear feasibility problem.

Define $d_0(\hat q)=\|\hat q-e_0\|$ and $B_r(e_0)=\{\hat q\in\bar{\mathcal O}_{d}|d_0(\hat q)<r\}$.
Set $\mathcal D:=\mathcal S(\mathcal H)\times\mathcal O_{d}$ and 
\(
\mathcal M_{\delta,r}
=\{(\rho,\hat q)\in\mathcal D|
\Tr(\Pi_0\rho)<\delta,\ \hat q_0>1-r\}.
\)
Define
\begin{equation}\label{eq:log_exit_function}
 W_\beta(\hat q):=-\sum_{n=1}^d\beta_n\log\hat q_n\geq0.
\end{equation}

\begin{lemma}
\label{Lem:Instability}
Suppose \textbf{A4}(i), \textbf{A5}, and
\textbf{A7} hold. There exist $\delta,r\in(0,1)$
and $c,C>0$ such that,  for almost every $t\geq0$,
\[
\mathscr L_tW_\beta\leq-c
\quad\text{on }\mathcal M_{\delta,r},
\quad
\mathscr L_tW_\beta\leq C
\quad\text{on }\mathcal D.
\]
Moreover, for every
$z\in\mathcal M_{\delta,r}$, $\mathbb E_z(\tau_{\mathcal M_{\delta,r}^c})\leq{W_\beta(\hat q(0))}/{c}$ where $\tau_{\mathcal M_{\delta,r}^c}
:=\inf\{t\geq0|Z(t)\notin\mathcal M_{\delta,r}\}$.
\end{lemma}
\proof
The infinitesimal generator of $W_\beta$ is given by
\begin{equation}\label{eq:log_exit_generator}
 \mathscr L_tW_\beta
=-u(\hat q)\sum_{n=1}^d
\beta_n\frac{(\Gamma\hat q)_n}{\hat q_n}
+\mathcal B_t(\rho,\hat q),
\end{equation}
where
\[
\mathcal B_t(\rho,\hat q):=\textstyle 2\sum_{n,k}
\beta_n\sqrt{\hat\theta_k}\Phi_n^k(\hat q) \bigl[\sqrt{\hat\theta_k}\Phi_n^k(\hat q)
-\mathfrak T_k(t,\rho,\hat q)\bigr].
\]
Since $\Gamma_{nj}\geq0$ for $n\ne j$
and $u\geq0$,
\[
-u(\hat q)\sum_{n=1}^d
\beta_n\frac{(\Gamma\hat q)_n}{\hat q_n}
\leq
u(\hat q)\sum_{n=1}^d
\beta_n(-\Gamma_{nn})
\longrightarrow0
\]
as $\hat q\to e_0$ since $u(e_0)=0$. For every $\rho\in\mathcal I_0$, we have
\[
\mathcal B_t(\rho,e_0)= -\sum_{j=1}^d p_j
  \sum_{n=1}^d\beta_nG_{nj}(\chi(t))\leq-\kappa_R,
\]
where $\sum_{j=1}^dp_j=1$.
By \textbf{A7}, this bound is uniform over $ [\underline{\chi}_k,\bar{\chi}_k]$.
By the uniform continuity on the compact state space
and parameter ranges, together with
$u(e_0)=0$, there exist
$\delta,r\in(0,1)$ such that $\mathscr L_tW_\beta\leq-\kappa_R/2$ on $\mathcal M_{\delta,r}$.
Therefore, $\mathscr L_tW_\beta\leq C$ on $\mathcal D$ for some constant $C>0$.
Define
\(
\sigma_j :=\inf\left\{t\geq0|\textstyle 
\min_{0\leq n\leq d}\hat q_n(t)\leq j^{-1}
\right\}.
\)
Since $W_\beta\geq0$,  by It\^o's formula, for sufficiently large $j$, we have
\(
 \mathbb E_z[t\wedge\tau_{\mathcal M_{\delta,r}^c}\wedge\sigma_j] \leq 2W_\beta(\hat q)/\kappa_R.
\)
By Theorem~\ref{Prop:Invariance_q}, $\sigma_j\uparrow\infty$ almost surely. Letting $j\to\infty$ and then
$t\to\infty$, the conclusion follows from monotone convergence.
\hfill$\square$

\subsubsection{Robust noncancellation}
Assumption~\textbf{A6} characterizes reachability
under the control Hamiltonian. However,
the remaining drift may cancel the
control-induced couplings.  Thus, we impose a structural condition ensuring that the reachability mechanism persists under perturbations.

Let $P_\ell$ denote the orthogonal projection onto $\mathcal R_\ell$ which is defined by~\eqref{eq:reach_sequence}. By Proposition~\ref{prop:joint_invariance} and its proof, we have $P_\ell\Pi_n=\Pi_nP_\ell$.
Write $K(t)=H_0(t)+\tilde H_0(t)$ and $Q_C(t)=\sum_rC_r(t)^*C_r(t).$ For brevity, define
\begin{align*}
 B_{\ell,n}&:=\Pi_n(\mathbf I-P_\ell)H_1P_\ell,\quad  D_{\ell,n}(t):=\Pi_n(\mathbf I-P_\ell)
       \left(K(t)+\tfrac{\mathfrak i}{2}Q_C(t)\right)P_\ell.
\end{align*}
The two maps have domain $\mathcal R_\ell$ and range in $\mathcal H_n\cap\mathcal R_\ell^\perp$.
\begin{description}
\item[\textbf{A8:}] For every $0\leq\ell<J$, $0\leq n\leq d$,
and almost every $t\geq0$, $D_{\ell,n}(t)B_{\ell,n}^* +B_{\ell,n}D_{\ell,n}(t)^*\geq 0.$
\end{description}
Indeed, \textbf{A8} implies, for every $a>0$,
\begin{equation}\label{eq:cross_gain_bound}
(D_{\ell,n}+aB_{\ell,n})  (D_{\ell,n}+aB_{\ell,n})^*\geq a^2B_{\ell,n}B_{\ell,n}^*.
\end{equation}
Thus, we have $\ker(D_{\ell,n}+aB_{\ell,n})^*\subseteq\ker B_{\ell,n}^*.$
This is the kernel-propagation property
required in the non-persistence argument
below.

\begin{proposition}
\label{prop:robust_certificates}
Under \textbf{A1} and \textbf{A2}, each of the following implies \textbf{A8}:
\begin{enumerate}[(i)]
\item $\mathcal R_1=\mathcal H$.
\item $[K(t),P_\ell]=[Q_C(t),P_\ell]=0$ for every $\ell<J$ a.e.
\item \textbf{A-qsr} holds and $ \Pi_nH_1\Pi_n=h_n\Pi_n$ with $ h_n\in\mathbb R$ for all $0\leq n\leq d$.
\end{enumerate}
\end{proposition}
\proof
By \textbf{A1} and \textbf{A2}, we deduce $\Pi_0^{\perp}K\Pi_0=0$ and $\Pi_0^{\perp}Q_C\Pi_0=\sum_rC_{r,P}^*C_{r,S}=0$.
If $\mathcal R_1=\mathcal H$, only $\ell=0$ occurs and all $D_{0,n}=0$. In case~(ii), all $D_{\ell,n}=0$ directly.
For~(iii), \textbf{A-qsr} implies that $K$ and $Q_C$ commute with
all $\Pi_n$. Since $P_\ell\Pi_n=\Pi_nP_\ell$, $D_{\ell,n}=D_{\ell,n}\Pi_n$ and $\Pi_nB_{\ell,n}^*
 =P_\ell\Pi_nH_1\Pi_n(\mathbf I-P_\ell)=0.$ Hence, $D_{\ell,n}B_{\ell,n}^*=0$.
\hfill$\square$


\begin{remark}
Assumption~\textbf{A8} provides a uniform
noncancellation certificate for admissible
time-varying perturbations.
For a real-analytic matrix family on a connected open
parameter domain, a rank lower bound verified at one
parameter value holds outside a Lebesgue-null set;
see~\cite{liang2025exploring}.
Such generic rank properties do not imply
\textbf{A8} or uniform closed-loop
non-persistence. An almost-everywhere
stabilization result would require
a separate recurrence analysis.
\end{remark}

For the recurrence analysis, let
$\mathfrak U$ denote the class of
admissible deterministic measurable
coefficient paths satisfying
\textbf{A1}, \textbf{A2}, and
\textbf{A8}, together with the common
bounds specified in
Section~\ref{Sec:SysDes}.
The fixed data are
$H_1,L_k,\Pi_n,\Gamma,u,
\hat\theta_k$, and all paths share
the certificate $(\beta,\kappa_R)$
in \textbf{A7}.
The class $\mathfrak U$ is closed
under nonnegative time translations.
We denote by $\mathbb P_z^{\mathfrak a}$
and $\mathbb E_z^{\mathfrak a}$
the probability measure and expectation of the
coupled process initialized at $z$.
All recurrence estimates below are
uniform over $\mathfrak a\in\mathfrak U$.

\subsubsection{Deterministic controlled equations and target access}
To establish target accessibility, we consider
the deterministic controlled equations associated
with the support of the coupled stochastic system~\eqref{Eq:SME} and~\eqref{Eq:SDE}.
Following~\cite{stroock1972support}, these equations
take the form
\begin{align}
\dot\rho_v(t)& =\tilde{\mathcal L}^u_{\gamma,\eta}(t,\rho_v(t))
 +\mathcal P(t,\rho_v(t))+\sum_k\sqrt{\theta_k(t)}\mathcal G_{L_k}(\rho_v(t))V_k(t),
 \label{Eq:ODE}\\
\dot{\hat q}_{v,n}(t)& =\mathfrak f_n^u(\hat q_v(t))+\Delta_n(\hat q_v(t))+2\hat q_{v,n}(t)\sum_k\sqrt{\hat\theta_k}
          \Phi_n^k(\hat q_v(t))V_k(t),\label{Eq:ODE_F}
\end{align}
where $u=u(\hat q_v)$, $v\in L^2([0,T];\mathbb R^m)$,
$V_k=v_k+\sqrt{\theta_k}\Tr((L_k+L_k^*)\rho_v)$, and
$(\rho_v(0),\hat q_v(0))=(\rho(0),\hat q(0))$.
Here
\begin{align*}
\small
&\tilde{\mathcal L}^u_{\gamma,\eta}(t,\rho)= -\mathfrak i[H_0(t)+uH_1,\rho]+\sum_k\frac{\gamma_k(t)}2\big[2(1-\eta_k(t))L_k\rho L_k^*-(L_k^*L_k+\eta_k(t)L_k^2)\rho\\
 &~~~~~~~~~~~~~~~~-\rho(L_k^*L_k+\eta_k(t)(L_k^*)^2)+\eta_k(t)\Tr((L_k+L_k^*)^2\rho)\rho\big],\\
&\mathfrak f_n^u(\hat q)= u\sum_j\Gamma_{n,j}\hat q_j-4\hat q_n\sum_k\hat\theta_k\Phi_n^k(\hat q)\Lambda_k(\hat q),\\
&\Delta_n(\hat q)= 2\hat q_n\sum_k\hat\theta_k\Big[\sum_j\hat q_j \Re\{l_{k,j}\}\Phi_j^k(\hat q)-(\Phi_n^k(\hat q))^2\Big].
\end{align*}
Lemma~\ref{lem:measurable_support} in
Appendix~\ref{App:support} establishes the
support property for these equations with
bounded measurable time-dependent coefficients.
Since $\mathcal X$ is closed and invariant
under the stochastic dynamics, the support
property (Lemma~\ref{lem:measurable_support}) also implies invariance of
$\mathcal X$ for the solution $Z_v(t):=(\rho_v(t),\hat q_v(t))$ of controlled equations~\eqref{Eq:ODE}--\eqref{Eq:ODE_F}.

The following lemma shows that the prediction
dynamics ensure strict positivity of the target
coordinate, even when its initial value is zero.

\begin{lemma}
\label{lem:estimator_recovery}
Assume that \textbf{A4} and \textbf{A5} hold. For the deterministic coupled system~\eqref{Eq:ODE}--\eqref{Eq:ODE_F} with $(\rho_v(0),\hat{q}_v(0))\in\mathcal X$, $\hat q_{v,0}(t)>0$ for all $t>0$. Moreover, if $\hat q_v(0)\ne e_0$, then $u(\hat q_v(t))>0$ for all $t\geq 0$.
\end{lemma}

\proof
Rewrite~\eqref{Eq:ODE_F} as
\[
\textstyle \dot{\hat q}_{v,i}
=a_i(t)\hat q_{v,i}
+u(\hat q_v)\sum_{j\ne i}
\Gamma_{ij}\hat q_{v,j},
\]
where $a_i$ is locally integrable.
Set
\[
E_i(t)=\exp\big(\int_0^t a_i(s)ds\big).
\]
By the variation-of-constants formula, we have
\begin{align}\small
\hat q_{v,i}(t)=&E_i(t)\Bigg[\hat q_{v,i}(0)+\int_0^t E_i(s)^{-1}u(\hat q_v(s))
\sum_{j\ne i}\Gamma_{ij}
\hat q_{v,j}(s)ds
\Bigg].\label{eq:coordinate_variation}
\end{align}
Since $\Gamma_{ij}\geq0$ for $i\ne j$,
every initially positive coordinate remains
strictly positive.
If $\hat q_v(0)=e_0$, then $\hat q_v(t)=e_0$ for all $t\geq0$.
Otherwise, some $j_0>0$ satisfies
$\hat q_{v,j_0}(0)>0$. Thus, $\hat q_v(t)\ne e_0$ for all $t\geq 0$.
By \textbf{A5}, we have $u(\hat q_v(t))>0$.

By \textbf{A4}(ii), there exists a directed path $j_0\to j_1\to\cdots\to j_\ell=0$ and $\Gamma_{j_{r+1}j_r}>0.$
If $\hat q_{v,j_r}(t)>0$ for all $t>0$,
by \eqref{eq:coordinate_variation}, we deduce
\[
\begin{aligned}\small
&\hat q_{v,j_{r+1}}(t) \geq E_{j_{r+1}}(t)\Gamma_{j_{r+1}j_r}\int_0^t E_{j_{r+1}}(s)^{-1}
u(\hat q_v(s))\hat q_{v,j_r}(s)\,ds
>0, \quad \forall t>0.
\end{aligned}
\]
Thus, we conclude $\hat q_{v,0}(t)>0$ for all $t>0$ by induction.
\hfill$\square$

We next provide the sufficient condition to exclude persistence of the trajectory $\rho_{v}(t)\in\mathcal I_0$ under positive feedback.

\begin{lemma}
\label{Lem:non_invariance}
Assume that \textbf{A1}-\textbf{A6} and
\textbf{A8} hold, and $\bar\eta_k<1$ for all $k$.
For any $0\leq a<b\leq T$, if $u(\hat q_v(t))>0$ for almost every $t\in [a,b]$, then $\sup_{t\in [a,b]}\Tr(\Pi_0\rho_v(t))>0.$
\end{lemma}

\proof
Suppose that $\rho_v(t)\in\mathcal I_0$  for all $t\in[a,b]$. Then, we have $\rho_v(t)\Pi_0=0$ for all $t\in[a,b]$, which means $\mathcal R_0 \subseteq \ker\rho_v(t)$ for all $t\in[a,b]$. Next, we show that each $\mathcal R_\ell \subseteq \ker\rho_v(t)$ for all $\ell$.

For any $w\in\ker\rho_v(t)$, $f(x)=w^*\rho_v(x)w\geq 0$ and $f(t)=0$, then $f(x)$ takes the minimum at $x=t$ and $\frac{d}{dx}f(x)|_{x=t}=0$. It implies
\begin{align*}
0=w^*\dot\rho_v(t)w
= \sum_k(\gamma_k\!-\!\theta_k)\|\rho_v^{1/2}L_k^*w\|^2\!+\!\sum_r\!\|\rho_v^{1/2}C_r^*w\|^2.
\end{align*}
Since $\gamma_k-\theta_k\geq\underline\gamma_k(1-\bar\eta_k)>0$ and $\ker\rho^{1/2}=\ker\rho$, it follows that $L_k^*\ker\rho_v(t)\subseteq\ker\rho_v(t)$ and $\rho_v(t)C_r(t)^*w=0.$
By \textbf{A3}, the vectors
$(l_{1,n},\ldots,l_{m,n})$ are pairwise
distinct. Thus, there exist $a_1,\ldots,a_m\in\mathbb C$ such that $\lambda_n:=\sum_{k=1}^m
a_k\overline{l_{k,n}},$ for all $n=0,\ldots,d$ are pairwise distinct. Define $M=\sum_{k=1}^m a_kL_k^*$. By Lagrange interpolation, we have
$M=\sum_{n=0}^d\lambda_n\Pi_n$ and
\[
\Pi_n=
\prod_{0\leq j\leq d, j\ne n}
\frac{M-\lambda_j\mathbf I}
{\lambda_n-\lambda_j}.
\]
Since $\ker\rho_v(t)$ is invariant
under each $L_k^*$, it is invariant
under $M$ and hence under every
$\Pi_n$. Thus, it is invariant under each $L_k$.

Assume that $\mathcal R_\ell\subseteq\ker\rho_v(t)$ for all $t\in[a,b]$. Then, we have $\rho_v(t)P_\ell=0$, which implies $\dot\rho_v(t)P_\ell=0$ almost everywhere on $[a,b]$.
By the preceding kernel invariance, for almost every $t\in[a,b]$,
\(
\rho_v L_kP_\ell =\rho_v L_k^*P_\ell=\rho_v C_r^*P_\ell=0.
\)
We deduce
\begin{align*}
0=\dot\rho_vP_\ell
&=\mathfrak i\rho_v(K+uH_1)P_\ell
-\rho_vQ_CP_\ell/2=\mathfrak i\rho_v
\big(K+\mathfrak i Q_C/2+uH_1\big)P_\ell.
\end{align*}
Since
$\Pi_n\ker\rho_v\subseteq\ker\rho_v$
and $[P_\ell,\Pi_n]=0$, we obtain, for all $n=0,\ldots,d$,
\begin{align*}
0
&=\rho_v\Pi_n
\big(K+\mathfrak i Q_C/2+uH_1\big) P_\ell=\rho_v\Pi_n(\mathbf I-P_\ell)\big(K+\mathfrak i Q_C/2+uH_1\big) P_\ell=\rho_v(D_{\ell,n}+uB_{\ell,n}).
\end{align*}
Multiplying by $B_{\ell,n}^*$ and taking
the real part of the trace yields
\begin{align*}
0=&\Re\big[ \Tr\bigl(\rho_v(D_{\ell,n}+uB_{\ell,n})B_{\ell,n}^*\bigr)\big]=u\Tr(\rho_vB_{\ell,n}B_{\ell,n}^*)+\Tr\bigl(\rho_v(D_{\ell,n}B_{\ell,n}^*+B_{\ell,n}D_{\ell,n}^*)\bigr)/2.
\end{align*}
By \textbf{A8}, $D_{\ell,n}B_{\ell,n}^*+B_{\ell,n}D_{\ell,n}^*\geq 0$. Moreover, we have
\[
\Tr(\rho_vB_{\ell,n}B_{\ell,n}^*)=\|\rho_v^{1/2}B_{\ell,n}\|_{\mathrm{HS}}^2\geq0.
\]
Since $u>0$ almost everywhere on $[a,b]$, $\|\rho_v^{1/2}B_{\ell,n}\|_{\mathrm{HS}}^2=0$ which implies $\rho_vB_{\ell,n}=0$.
Using $[P_\ell,\Pi_n]=0$ and
$\rho_vP_\ell=0$, we obtain
\begin{align*}
\rho_v\Pi_nH_1P_\ell &=\rho_v\Pi_n(\mathbf I-P_\ell)
H_1P_\ell+\rho_v\Pi_n P_\ell H_1 P_\ell =\rho_vB_{\ell,n}+\rho_vP_\ell\Pi_nH_1P_\ell=0.
\end{align*}
Therefore, we have
\[
\mathcal R_{\ell+1}=\textstyle \mathcal R_\ell+ \sum_{n=0}^d\Pi_nH_1\mathcal R_\ell \subseteq\ker\rho_v(t), \,\text{for a.e. }t\in[a,b].
\]
By induction, \textbf{A6} and the continuity of $\rho_v(t)$, we deduce
\[
\mathcal H=\mathcal R_J
\subseteq\ker\rho_v(t),
\quad \forall t\in[a,b],
\]
which implies $\rho_v(t)=0$ contradicting $\Tr(\rho_v(t))=1$.
\hfill$\square$

The non-persistence property yields the following
uniform preparation estimate.

\begin{lemma}
\label{lem:uniform_preparation}
Assume \textbf{A1}--\textbf{A6}, \textbf{A8}, and
$\bar\eta_k<1$ for all $k$.
For every $r\in(0,1)$, there exist $u_*,a_q,a_\rho>0$ such that the solution of~\eqref{Eq:ODE}--\eqref{Eq:ODE_F} with
$V_k=0$ on $[0,1]$ and $1-\hat q_{v,0}(0)\geq r$ satisfies 
$$
\inf_{0\leq t\leq1}u(\hat q_v(t)) \geq u_*, \quad \inf_{1/2\leq t\leq1}\hat q_{v,0}(t)\geq a_q, \quad \max_{1/2\leq t\leq1}\Tr(\Pi_0\rho_v(t))\geq a_\rho,
$$
where these constants are uniform over $\mathfrak a\in\mathfrak U$ and $Z_v(0)\in\mathcal X$.
\end{lemma}

\proof
Fix $\mathfrak a\in\mathfrak U$ and $Z_v(0)\in\mathcal X$, consider the  trajectory with $V_k=0$ on $[0,1]$.
Due to the compactness, there exists $C_1>0$ such that $a_i(t)\geq -C_1$. Then, for every $i$,
\[
\textstyle \dot{\hat q}_{v,i} \geq -C_1\hat q_{v,i} +u(\hat q_v)\sum_{j\ne i}\Gamma_{ij}\hat q_{v,j}, \quad\text{a.e. on }[0,1].
\]
Summing over $i>0$, we obtain
\[
1-\hat q_{v,0}(t)\geq(1-\hat q_{v,0}(0))e^{-C_1t} \geq r e^{-C_1t}.
\]
Thus, by \textbf{A5}, for all $t\in[0,1]$,
\[
u(\hat q_v(t)) \geq \min\{u(\hat q) \mid \hat q\in\bar{\mathcal O}_{d}, \, 1-\hat q_0\geq r e^{-C_1}\} =:u_*>0.
\]
By the variation-of-constants formula, for all $t\in[0,1]$ and for any $i\neq j$, we obtain 
\begin{equation}\label{Eq:variation of constants ODE}
\hat q_{v,i}(t) \geq  e^{-C_1t}\hat q_{v,i}(0) +u_* \Gamma_{ij}\int^t_0 e^{-C_1(t-s)}\hat q_{v,j}(s)ds,
\end{equation}
where we have used the property $\Gamma_{ij}\geq 0$ and $\hat{q}_{v,j}\geq 0$.
Choose $j_0>0$ with $\hat q_{v,j_0}(0)\geq r/d$, then $\hat q_{v,j_0}(t) \geq  \frac {r}{d} e^{-C_1t}$ due to~\eqref{Eq:variation of constants ODE}. By \textbf{A4}(ii), for each $j\in[d]$, there is a simple directed path $j_0\to j_1\to\cdots\to j_{\ell_j}=0$, where $\ell_j\in [d]$ and $\Gamma_{j_{h+1}j_h}>0$. For the first iteration $j_0\to j_1$, by~\eqref{Eq:variation of constants ODE}, we obtain
$
\hat q_{v,j_1}(t) \geq \frac{r}{d} e^{-C_1 t}  (u_* t)  \Gamma_{j_1 j_0}.
$
Then, by iteration, we deduce
\begin{equation}\label{Eq: ineq q_v0}
\hat q_{v,0}(t)\geq   \frac{r(u_*t)^{\ell_j}e^{-C_1t}}{(\ell_j!)d}\prod_{h=0}^{\ell_j-1}\Gamma_{j_{h+1}j_h}, \quad \forall \, t\in(0,1].
\end{equation}
Taking the minimum over the finitely many paths and $t\in[1/2,1]$ yields
$a_q>0$ such that $\hat q_{v,0}(t)\geq a_q$ on $[1/2,1]$. Set 
$$
a_q:= \frac {re^{-C_1}}{d}
\min_{j\in[d]}\left\{
\frac{(u_*/2)^{\ell_j}}{\ell_j!}
\prod_{h=0}^{\ell_j-1}\Gamma_{j_{h+1}j_h}
\right\}>0.
$$ 
which is independent of the initial state and admissible coefficients, then we have $\hat q_{v,0}(t)\geq a_q$ for all $t\in[1/2,1]$.

We next establish the uniform bound.
Set $K=H_0+\tilde H_0$,
$\mathcal J_t(X)=\sum_r C_r(t)XC_r(t)^*$,
and $Q_C=\mathcal J^*(\mathbf I)$.
For $V=0$, the controlled drift is affine in
$\xi=(K,\mathcal J,\gamma,\theta)$.
Let $\mathscr C$ be the set of such tuples satisfying
\begin{align*}
&K=K^*,~~\|K\|\leq M_K,~~\operatorname{Ch}(\mathcal J)\geq 0,~~\Tr[\operatorname{Ch}(\mathcal J)]\leq M_J,\\
&\Pi_0^{\perp}K\Pi_0=\Pi_0^\perp\mathcal J(\Pi_0)\Pi_R=\Pi_0\mathcal J^*(\mathbf I)\Pi_0^{\perp}=0,\\
&\underline\gamma_k\leq\gamma_k\leq\bar\gamma_k,\quad
\underline\eta_k\gamma_k \leq\theta_k\leq\bar\eta_k\gamma_k, \quad \forall \, k\in[m],
\end{align*}
together with \textbf{A8}, where $\operatorname{Ch}(\mathcal J)$ denotes
the Choi matrix, $M_K=M_0+M_{\mathrm p}$,
and $M_J=\sum_r M_{C,r}^2$.
All constraints are closed and convex.
Thus, $\mathscr C$ is compact and convex
and contains the admissible coefficient values.

Suppose that no uniform $a_\rho>0$ exists. Then, there are admissible  paths $\xi^{(h)}$ and $Z^{(h)}_v(0)\in\mathcal X$ such that $\lim_{h\to \infty}\max_{t\in[1/2,1]} \Tr(\Pi_0\rho_v^{(h)}(t))=0$.
Since $\xi^{(h)}$ is uniformly bounded in $L^\infty([0,1])$ and the corresponding trajectories are uniformly bounded and
equi-Lipschitz, by Banach-Alaoglu theorem and Arzel\`a-Ascoli theorem, and the separability of $L^1([0,1])$,
we can extract a
subsequence such that
\[
\xi^{(h)}\rightharpoonup^*\xi \text{ in }L^\infty([0,1]),
\quad Z_v^{(h)}\to Z_v \text{ uniformly on }[0,1].
\]
To pass to the limit, write the drift of the coupled equation~\eqref{Eq:ODE}--\eqref{Eq:ODE_F} in
real coordinates as $F(Z_v,\xi)=F_0(Z_v)+\sum_{j=1}^M\xi_jF_j(Z_v)$, where $F_j$ are Lipschitz on $\mathcal X$.
Then, for every $t\in[0,1]$,
\begin{align*}\small
&\int_0^t\xi_j^{(h)}(s)
\bigl[F_j(Z_v^{(h)}(s))-F_j(Z_v(s))\bigr]\,ds
\longrightarrow0,\\
&\int_0^t
\bigl[\xi_j^{(h)}(s)-\xi_j(s)\bigr]
F_j(Z_v(s))\,ds
\longrightarrow0,
\end{align*}
where the first convergence follows from the
uniform convergence of trajectories, and the second follows from weak-star
convergence. Hence, $Z_v$ solves the controlled equation
with coefficients $\xi$.

Due to the compactness and convexity of $\mathscr C$, $\xi(t)\in\mathscr C$ almost everywhere.
In particular, $\mathcal J_t$ admits a bounded measurable Kraus representation, the target-preserving constraints and
\textbf{A8} remain valid, and $\gamma_k-\theta_k \geq\underline\gamma_k(1-\bar\eta_k)>0.$ Hence, the kernel argument of Lemma~\ref{Lem:non_invariance} applies to the limiting system. However, uniform convergence gives $\Tr(\Pi_0\rho_v(t))=0$ for all $t\in[1/2,1]$, and $u(\hat q_v(t))\geq u_*>0$ for all $t\in[0,1]$, which contradicts Lemma~\ref{Lem:non_invariance}. Thus, $a_\rho>0$ exists.
\hfill$\square$

The uniform preparation and support properties,
together with the logarithmic exit estimate,
yield the following recurrence result.

For all $k\in[m]$, set
$$
 \Psi_0^k(\rho):=\Re\{l_{k,0}\}-\sum_{j=0}^d\Re\{l_{k,j}\}\Tr(\Pi_j\rho),
$$
and define
\begin{align*}
\bar{\mathfrak{c}}_k:=\Re\{l_{k,0}\}\!-\!\min_{n\in[d]}\!\Re\{l_{k,n}\}, \ \underline{\mathfrak{c}}_k:=\Re\{l_{k,0}\}\!-\!\max_{n\in[d]}\!\Re\{l_{k,n}\}.
\end{align*}
Impose the following assumption on measurement operators:
\begin{description}
    \item[\textbf{A9:}] For each $k\in[m]$, $\bar{\mathfrak{c}}_k\leq 0$ or $\underline{\mathfrak{c}}_k\geq 0$.
\end{description}
Assumption~\textbf{A9} requires the target measurement
value to be either the minimum or the maximum
among all sector values in each channel.
This ordering prevents cancellation of
target--non-target measurement contrasts
within individual channels. Choose $s_k\in\{-1,1\}$ so that
$s_k(\Re l_{k,0}-\Re l_{k,n})\ge0$ for every $n>0$. By \textbf{A3} and \textbf{A9}, we have
\begin{align*}
\sum_k s_k\Psi_0^k(\rho)=\sum_{n>0}p_n\sum_k|\Re l_{k,0}-\Re l_{k,n}|\ge (1-p_0)\min_{n>0}\sum_k|\Re l_{k,0}-\Re l_{k,n}|>0
\end{align*}
whenever $p_0<1$. Thus,  the channels jointly separate the target from every
non-target population mixture. 

\begin{proposition}
\label{Prop:Recurrence}
Assume \textbf{A1}--\textbf{A9} and
$\bar\eta_k<1$ for all $k\in[m]$.
For $\zeta>0$, let
$\mathcal N_\zeta
=\mathbf B_\zeta(\mathcal H_0)\times B_\zeta(e_0)$
and
$\tau_\zeta=\inf\{t\geq0|Z(t)\in\mathcal N_\zeta\}$. Then, there exist $T,c,C>0$ and $p_*\in(0,1]$ independent of $\mathfrak a\in\mathfrak U$ and $z=(\rho(0),\hat q(0))\in\mathcal D$, such that
\begin{equation}\label{eq:uniform_mean_statement}
\mathbb E_z^{\mathfrak a}[\tau_\zeta]
\leq
\frac{W_\beta(\hat q(0))}{c}
+\frac{C+c}{c}\frac{T}{p_*}
<\infty.
\end{equation}
In particular, $\mathbb P_z^{\mathfrak a}(\tau_\zeta<\infty)=1$.
\end{proposition}

\proof
Let $\delta,r,c,C$ be given by
Lemma~\ref{Lem:Instability}, and set
$\mathcal K:=\{(\rho,\hat q)\in\mathcal X\setminus \mathcal{N}_{\zeta}|\Tr(\Pi_0\rho)\geq \delta \text{ or }\hat{q}_0\leq 1-r\}$
which is compact.
If $\mathcal K=\varnothing$, the conclusion follows directly from the negative drift
estimate below. Hence, assume $\mathcal K\neq \varnothing$.

Fix $\mathfrak a\in\mathfrak U$ and $z\in\mathcal K$, and write
$p_{v,0}(t)=\Tr(\Pi_0\rho_v(t))$. If $\hat q_{v,0}(0)>1-r$, then $p_{v,0}(0)\geq\delta$, and  take $t_1=0$. Otherwise, Lemma~\ref{lem:uniform_preparation} with $V_k=0$ gives some $t_1\in[1/2,1]$ such that $p_{v,0}(t_1)\geq a_\rho$ and $\hat q_{v,0}(t_1)\geq a_q$.
Thus, in both cases, $t_1\in[0,1]$ and $p_{v,0}(t_1),\hat q_{v,0}(t_1)\geq2a$ where $a=\frac12\min\{\delta,1-r,a_\rho,a_q\}>0$.

By \textbf{A9}, choose $s_k\in\{-1,1\}$
such that
$s_k(\Re\{l_{k,0}\}-\Re\{l_{k,j}\})\geq0$
for all $j>0$. In particular,
\begin{align}
\underline{\mathfrak c}_k(1-p_{v,0}) &\leq\Psi_0^k(\rho_v) \leq\bar{\mathfrak c}_k(1-p_{v,0}),\label{Eq:Psi}\\
\underline{\mathfrak c}_k(1-\hat q_{v,0}) &\leq\Phi_0^k(\hat q_v) \leq\bar{\mathfrak c}_k(1-\hat q_{v,0}).\label{Eq:Phi}
\end{align}
By \textbf{A3}, we deduce
\begin{equation*}
\small
\mu:=
\min_{j>0}\textstyle \sum_{k}
\min\{\sqrt{\underline\theta_k},
       \sqrt{\hat\theta_k}\}
s_k(\Re\{l_{k,0}\}-\Re\{l_{k,j}\})>0.
\end{equation*}
Thus, by taking $V_k=Ms_k$, there exists a constant $C_0>0$ which is independent of
$\mathfrak a,z$ and $M$ such that
\[
\dot x\geq-C_0+2M\mu x(1-x),
\quad x\in\{p_{v,0},\hat q_{v,0}\}.
\]
Choose $\epsilon\in(0,1/2)$ such that $a<1-\epsilon$ and $\sqrt{2\epsilon}<\zeta/2$. Taking $M>0$ sufficiently large that $2M\mu a\epsilon\geq C_0+1$,
we have $\dot x\geq1$ whenever $x\in[a,1-\epsilon]$. Therefore, we have
\[
p_{v,0}(t),\hat q_{v,0}(t)\geq1-\epsilon, \quad \forall \, t\in[t_1+1,2].
\]
At $T=2$, $\mathbf d_0(\rho_v(T))^2\leq2\epsilon$ and $d_0(\hat q(T))^2\leq2\epsilon$, which implies $Z_v(T)\in\mathcal N_{\zeta/2}$.
Then, by applying Lemma~\ref{lem:measurable_support} and the compactness of $\mathcal K$, there exists a constant $p_*$ such that, 
\begin{equation}\label{eq:uniform_entrance}
\inf_{\mathfrak a\in\mathfrak U,z\in\mathcal K} \mathbb P_z^{\mathfrak a}(\tau_\zeta\leq T)\geq p_*>0.
\end{equation}

Next, we estimate the occupation time. Fix $\mathfrak a\in\mathfrak U$ and
$z\in\mathcal D$. Since $\mathfrak U$ is invariant under time translations, by the Markov property and \eqref{eq:uniform_entrance}, on the event $\{t<\tau_\zeta,\ Z(t)\in\mathcal K\}$, we obtain $\mathbb P(\tau_\zeta\leq t+T\mid\mathcal F_t) \geq p_*$, which implies
\begin{align}
p_*\mathbb E\!\!\int_0^{\tau_\zeta} \mathds1_{\mathcal K}(Z(t))dt \leq \mathbb E\int_0^{\tau_\zeta}  \mathds1_{\mathcal K}(Z(t)) \mathds1_{\{\tau_\zeta \leq t+T\}} dt\leq T,
\label{eq:occupation_K}
\end{align}
where we used $\int_0^{\tau_\zeta} \mathds1_{\{\tau_\zeta\leq t+T\}}dt =(\tau_\zeta\wedge T)\mathds1_{\{\tau_\zeta<\infty\}} \leq T$.

By Lemma~\ref{Lem:Instability}, we have
$$\mathscr L_tW_\beta \leq-c+(C+c)\mathds1_{\mathcal K} \text{ on } \mathcal D\setminus\mathcal N_\zeta.$$
Let
$\sigma_j=\inf\{t\geq0|\min_n\hat q_n(t)\leq j^{-1}\}$.
Applying It\^o's formula up to
$t\wedge\tau_\zeta\wedge\sigma_j$,
and using $W_\beta\geq0$, we deduce
\[
\begin{aligned}
c\mathbb E[t\wedge\tau_\zeta\wedge\sigma_j]
&\leq W_\beta(\hat q)+(C+c)\mathbb E
\int_0^{t\wedge\tau_\zeta\wedge\sigma_j}
\mathds1_{\mathcal K}(Z(s))ds\leq W_\beta(\hat q)+(C+c)T/p_*.
\end{aligned}
\]
Since $\sigma_j\uparrow\infty$ almost surely
by Theorem~\ref{Prop:Invariance_q},
letting $j\to\infty$ and then
$t\to\infty$ proves
\eqref{eq:uniform_mean_statement}
by monotone convergence.
\hfill$\square$

\subsubsection{Exponential stability}

For $k\in[m]$ and $j\in[d]$, set 
$\bar{\mathfrak l}_k:=\max\{|\underline{\mathfrak c}_k|,
|\bar{\mathfrak c}_k|\}$ and
\(
d_{kj}:=\big|\Re\{l_{k,0}\}-\Re\{l_{k,j}\}\big|.
\) Define
\begin{align*}
&S_A:= \min_{\substack{w_j\geq0\\\sum_{j}w_j=1}}
\sum_{k}\hat\theta_k\min\{\underline\chi_k^2,1\}
\left(\sum_{j=1}^dw_jd_{kj}\right)^2,\\
&S_B:= \sum_k\hat\theta_k2\bar{\mathfrak l}_k|\Re\{l_{k,0}\}|
\max\{1-\underline\chi_k,\bar\chi_k-1\}.
\end{align*}
The next condition ensures that the measurement-induced
contraction dominates the effect of parameter mismatch.
\begin{description}
\item[\textbf{A10:}] $S_A>S_B$.
\end{description}
Under \textbf{A3} and \textbf{A9}, we have $S_A>0$.
The definition of $S_A$ accounts for the collective
separation of the target from non-target sectors across
measurement channels, without requiring
any single channel to distinguish all non-target sectors. The
constant $S_A$ is computable by a convex quadratic program over
the probability simplex.
Write $\Delta(\rho,\hat q):=\mathbf d_0(\rho)+\|\hat q-e_0\|$, with the convention $\log 0 =-\infty$. Under \textbf{A1}, \textbf{A2} and \textbf{A5}, the joint target $\mathcal{I}(\mathcal{H}_0)\times\{e_0\}$ is invariant.
\begin{theorem}
\label{Thm:GES Feedback}
Assume \textbf{A1}--\textbf{A10} and $\bar\eta_k<1$ for all $k$.
Then, $\mathcal I(\mathcal H_0)\times\{e_0\}$ is stable in
probability uniformly over $\mathfrak U$, that is, for every $\varepsilon>0$, $z=(\rho(0),\hat q(0))\in \mathcal X$
\[\small
\lim_{\delta\downarrow0}
\sup_{\mathfrak a\in\mathfrak U}
\sup_{\Delta(z)<\delta}\mathbb P_z^{\mathfrak a}
\Big(
\sup_{t\ge0}\Delta(Z(t))\ge\varepsilon
\Big)=0.
\]
Moreover, for every
$\mathfrak a\in\mathfrak U$ and $z\in\mathcal D$,
\begin{equation}\label{eq:global_sample_exponent}
\limsup_{t\to\infty}\frac1t\log\Delta(Z(t))
\leq -(S_A-S_B),\quad a.s.
\end{equation}
In particular, $\rho(t)$ converges almost surely exponentially to $\mathcal{I}(\mathcal{H}_0)$ with sample exponent at most $-(S_A-S_B)$.
\end{theorem}

\proof
Let $\mathsf V_0=1-p_0+1-\hat q_0$ and $V=\mathsf V_0^x$, $x\in(0,1)$. For $\mathsf V_0>0$, the infinitesimal generator is given by
\begin{equation}\label{eq:local_lyapunov_generator}
\mathscr L_tV=xV\left[\textstyle \frac{\mathfrak F}{\mathsf V_0}
-2(1-x)\frac{\sum_k\mathfrak G_k}{\mathsf V_0^2}
\right],
\end{equation}
where $\mathfrak G_k=\hat\theta_k\big[\chi_k(t)\Tr(\Pi_0\rho)\Psi_0^k(\rho)
+\hat q_0\Phi_0^k(\hat q)\big]^2$ and
\begin{align*}
\mathfrak F=& -u(\hat q)\Big[\sum_j\Gamma_{0j}\hat q_j
-\mathfrak i\Tr(\Pi_0[H_1,\rho])\Big] -2\hat q_0\sum_k\sqrt{\hat\theta_k} \Phi_0^k(\hat q)\mathfrak T_k
-\sum_r\Tr(C_{r,P}\rho_R C_{r,P}^*).
\end{align*}
The last term of $\mathfrak F$ is nonpositive due to
\textbf{A2}. By \textbf{A5}, \textbf{A9},
\eqref{Eq:Psi}--\eqref{Eq:Phi}, and
$\mathfrak T_k=2\sqrt{\hat\theta_k}
[(\chi_k-1)\Re\{l_{k,0}\}-\chi_k\Psi_0^k+\Phi_0^k]$,
there exists a $C_0>0$ such that, for
$0<\mathsf V_0<\min\{1,r_u\}$,
\begin{align}
&\textstyle {\mathfrak F}/{\mathsf V_0}\leq 2S_B+C_0(\mathsf V_0^\alpha+\mathsf V_0),
\label{eq:local_drift_ratio}\\
&\textstyle {\sum_k\mathfrak G_k}/{\mathsf V_0^2}
\geq S_A(1-\mathsf V_0)^2.
\label{eq:local_quadratic_bound}
\end{align}
where $\alpha,r_u>0$ are defined in \textbf{A5}.
To prove \eqref{eq:local_quadratic_bound}, choose the signs
$s_k\in\{-1,1\}$ as in \textbf{A9}. Then,
$s_k\Psi_0^k(\rho)=\sum_{j=1}^d d_{kj}p_j$ and
$s_k\Phi_0^k(\hat q)=\sum_{j=1}^d d_{kj}\hat q_j$.
Since $\chi_k(t)\geq\underline\chi_k$ and
$p_0,\hat q_0\geq1-\mathsf V_0$, the two terms inside
the square defining $\mathfrak G_k$ have the same sign, and
\[
 \sqrt{\mathfrak G_k}\geq
\sqrt{\hat\theta_k}\min\{\underline\chi_k,1\}
(1-\mathsf V_0)\sum_{j=1}^d d_{kj}(p_j+\hat q_j).
\]
For $\mathsf V_0>0$, set
$w_j:=(p_j+\hat q_j)/\mathsf V_0$, $j\in[d]$.
Then, $w_j\geq0$ and $\sum_jw_j=1$. Hence, we deduce
\[
 {\sum_k\mathfrak G_k}/{\mathsf V_0^2}
\geq(1-\mathsf V_0)^2
\sum_k\hat\theta_k\min\{\underline\chi_k^2,1\}
\big(\sum_{j=1}^dw_jd_{kj}\big)^2,
\]
which proves \eqref{eq:local_quadratic_bound}
by the definition of $S_A$.
Choose $0<x<1-S_B/S_A$ and put
$\lambda_x:=x[(1-x)S_A-S_B]>0$.
For sufficiently small $r\in(0,1)$, uniformly over $\mathfrak U$, we have
\begin{equation}\label{eq:local_stability_drift}
\mathscr L_tV\leq-\lambda_xV,
\quad \text{if }0<\mathsf V_0<r.
\end{equation}
By applying the standard localized Lyapunov argument, see \cite[Theorem 6.3]{liang2019exponential} and \cite[Theorem 4.2.2]{mao2007stochastic}, we can prove the stability in probability, uniformly over $\mathfrak U$.


Combining this estimate with Proposition~\ref{Prop:Recurrence}, the strong Markov
property, and the time-translation invariance of
$\mathfrak U$,  we apply the recurrence--stability argument of
\cite[Theorem~5.7]{khasminskii2011stochastic} to the joint target set, and conclude that $\mathsf V_0(Z(t))\to0$ almost surely for all $z\in\mathcal D$.



Next, we estimate the convergence rate. By Theorem~\ref{Prop:Invariance_q}, 
$\mathsf V_0(Z(t))>0$ for all $t>0$ almost surely. Then, by applying It\^o's formula, we have
\begin{align*}
&d\log\mathsf V_0=\left( \frac{\mathfrak F}{\mathsf V_0}-2\frac{\sum_k\mathfrak G_k}{\mathsf V_0^2}\right)dt+dM_t,\\
&dM_t= -\frac2{\mathsf V_0}\sum_k\sqrt{\hat\theta_k}
\big[\chi_k\Tr(\Pi_0\rho)\Psi_0^k +\hat q_0\Phi_0^k\big]dW_k.
\end{align*}
Since $|\Psi_0^k|\leq\bar{\mathfrak l}_k (1-\Tr(\Pi_0\rho))$ and
$|\Phi_0^k|\leq\bar{\mathfrak l}_k(1-\hat q_0)$, there exists a $C_M>0$ such that  $\langle M\rangle_t\leq C_Mt$, which implies $M_t/t\to0$ almost surely. By
\eqref{eq:local_drift_ratio}--\eqref{eq:local_quadratic_bound}
and $\mathsf V_0(Z(t))\to0$, 
\[
\limsup_{t\to\infty}\frac1t\log\mathsf V_0(Z(t))
\leq-2(S_A-S_B)\quad a.s.
\]
Since $\mathbf d_0(\rho)^2\leq2(1-\Tr(\Pi_0\rho))$
and $\|\hat q-e_0\|\leq\sqrt2(1-\hat q_0)$, we have $\Delta\leq\sqrt{2\mathsf V_0}+\sqrt2\mathsf V_0$, which yields \eqref{eq:global_sample_exponent}.\hfill$\square$

\begin{remark}
\label{rem:robustness_margins}
 \textbf{A10} provides a sufficient margin
for uncertainty in the measurement parameters.
At the matched reference
$\underline\chi_k=\bar\chi_k=1$, we have $S_B=0$ and
$S_A>0$ by \textbf{A3} and \textbf{A9}.
If \textbf{A7} also holds, continuity
ensures that both \textbf{A7} and \textbf{A10} remain
satisfied for sufficiently small parameter uncertainty.
For fixed nominal parameters, enlarging the admissible
uncertainty intervals cannot increase $S_A$ or decrease
$S_B$. Thus, it reduces or preserves the 
convergence rate bound $S_A-S_B$.
This provides a criterion for selecting nominal parameters
under measurement uncertainty; see also
\cite{liang2021robust}.


Parameter-estimation methods
\cite{liang2024parameter}
can guide the selection of nominal parameters,
subject to the model and identifiability
assumptions. The residual uncertainty and 
parameter drift must be covered by bounds satisfying
\textbf{A7} and \textbf{A10}.
Thus, exact parameter identification is unnecessary:
within the certified ranges and under the stated
structural hypotheses, almost-sure exponential
convergence is preserved.
\end{remark}

\section{Numerical example: A three-level system}
\label{sec:simulation}

We consider a three-level spin system under continuous
homodyne monitoring of $J_z$, with target
$\Pi_0=\operatorname{diag}(1,0,0)$. Set
$L=J_z=\operatorname{diag}(1,0,-1)$,
$H_0(t)=\omega(t)J_z$, and
\[
\textstyle H_1= \frac{1}{\sqrt2}\left[
\begin{smallmatrix}
0&-\mathfrak i&0\\
\mathfrak i&0&-\mathfrak i\\
0&\mathfrak i&0
\end{smallmatrix}\right],\quad
\tilde H_0(t)=\left[
\begin{smallmatrix}
0.2&0&0\\
0&0.1&h(t)\\
0&h(t)&0.4
\end{smallmatrix}\right],
\]
where $\omega(t)=1.5[1+0.2\sin(2\pi t/T)]$
and $h(t)=0.5+0.2\sin(2\pi t/T)$.
The time-varying measurement parameters are $\gamma(t)=\gamma_0[1+0.2\sin(2\pi t/T)]$ and $\eta(t)=\eta_0[1+0.1\operatorname{tri}(t/T)]$,
where $\gamma_0 = 1.6$, $\eta_0=0.4$ and $\operatorname{tri}(s)=2\arcsin(\sin(2\pi s))/\pi$.
Thus, $\bar\eta=0.44$ and $\theta(t)\in[0.4608,0.8448]$. The additional dissipative channels are
$C_1(t)=\sqrt{\gamma_\phi(t)}J_z$ and $C_2(t)=\sqrt{\gamma_\downarrow(t)}|1\rangle\langle2|$, which describe additional dephasing and an incoherent transition between non-target levels, where $\gamma_\phi(t)=1.5[1+0.2\cos(2\pi t/T)]$ and $\gamma_\downarrow(t)=1.5[1+0.1\min\{t,100\}/T]$.
For the reduced filter, set $\hat\eta=0.4$,
$\hat\gamma=1.5$, and $\hat\theta=0.6$, with
\[
\Gamma=\left[
\begin{smallmatrix}
-1&1&0\\
1&-2&1\\
0&1&-1
\end{smallmatrix}\right],
\quad
u(\hat q)=4(1-\hat q_0)^2.
\]

The operators satisfy \textbf{A1}--\textbf{A5}
and \textbf{A9}. In particular, the target is preserved,
whereas $\tilde H_0$ and $C_2$ induce coherent and
dissipative transitions between the non-target sectors.
The reachability sequence satisfies
$\mathcal R_0=\operatorname{span}\{|0\rangle\}$,
$\mathcal R_1=\operatorname{span}\{|0\rangle,|1\rangle\}$,
and $\mathcal R_2=\mathbb C^3$, verifying \textbf{A6}.
Moreover, $\Gamma$ is obtained from
Lemma~\ref{lem:hamiltonian_laplacian} with $\kappa=2$.
For the noncancellation condition \textbf{A8},
$Q_C=\gamma_\phi J_z^2+\gamma_\downarrow\Pi_2$
is diagonal, and the only nonzero higher-layer
cross-blocks are $B_{1,2}={\mathfrak i}|2\rangle\langle1|/{\sqrt2}$ and $D_{1,2}=h(t)|2\rangle\langle1|
=-\mathfrak i\sqrt2h(t)B_{1,2}.$
Hence,
$D_{\ell,n}B_{\ell,n}^*
+B_{\ell,n}D_{\ell,n}^*=0$
for every relevant $\ell,n$, and \textbf{A8} holds. 
For \textbf{A7}, choosing $\beta_1=\beta_2=1/2$, we have $\textstyle \sum_n\beta_nG_{n1}(\chi)=\hat\theta$ and $\sum_n\beta_nG_{n2}(\chi)=\hat\theta(1+6\chi)$,
then $\kappa_R=\hat\theta$ is admissible.
Finally, we have $S_A=0.4608$ and $S_B\simeq0.4478$, which implies $S_A-S_B\simeq0.0130>0$ verifying \textbf{A10}. By Theorem~\ref{Thm:GES Feedback}, we obtain
\[
\limsup_{t\to\infty}\frac1t\log \mathbf{d}_0(\rho(t)) \leq \limsup_{t\to\infty}\frac1t\log\Delta(Z(t))\leq-0.0130. 
\]

\begin{figure}[!t]
\centering
\includegraphics[width=.49\columnwidth]{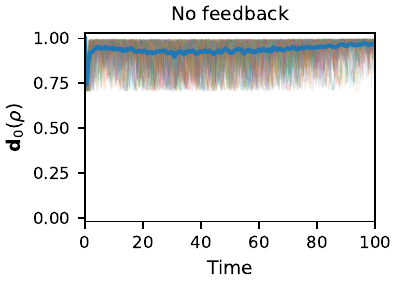}
\includegraphics[width=.49\columnwidth]{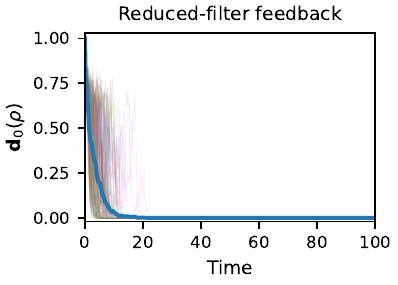}
\par
\includegraphics[width=.49\columnwidth]{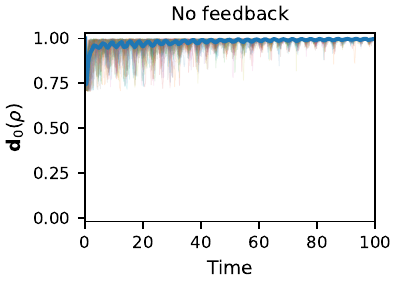}
\includegraphics[width=.49\columnwidth]{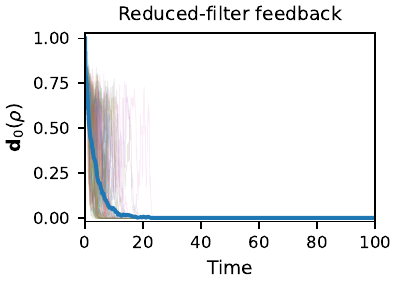}
\caption{Target-preserving perturbations:
$T=100\tau$ (top) and $T=2\tau$ (bottom).
Left: no feedback. Right: reduced-filter feedback.
Thin curves show $\mathbf d_0(\rho)$ for $100$ trajectories;
thick curves are empirical means.
Both rows use the same initial states, controller,
sample size, and time step.}
\label{Fig:Preserving}
\end{figure}

Let $\tau=1/(\eta_0\gamma_0)=1.5625$.
We consider slow ($T=100\tau$) and fast
($T=2\tau$) parameter variations, with
$\rho(0)=\Pi_2$ and $\hat q(0)=\mathbf1/3$.
Each case uses $100$ trajectories and time step $0.005$.
The physical state is simulated using Kraus-type scheme update~\cite{rouchon2015efficient},
while the reduced filter uses the same observation
record with fixed estimated parameters.
Fig.~\ref{Fig:Preserving} compares the
uncontrolled and feedback-controlled evolutions.
Without feedback, $\mathcal H_0^\perp$ is invariant
and $\Tr(\Pi_0\rho(t))=0$ for the chosen initial state.
The example illustrates robust target stabilization
under perturbations.

\begin{figure}[!t]
\centering
\includegraphics[width=.49\columnwidth]{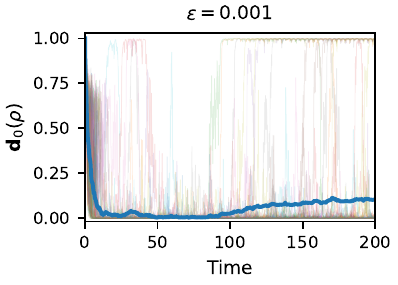}
\includegraphics[width=.49\columnwidth]{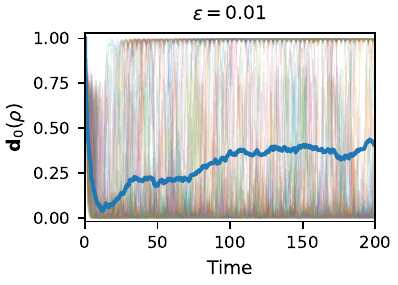}
\caption{Reduced-filter feedback under target leakage, violating \textbf{A2}, with $T=2\tau$.
Left: $\varepsilon=0.001$. Right: $\varepsilon=0.01$.
Thin curves show $\mathbf d_0(\rho)$ for $100$ trajectories;
the thick curve is their empirical mean.}
\label{Fig:Leakage}
\end{figure}

Next, we add the unobserved channel
$C_{\mathrm{leak}}=\sqrt{\varepsilon}|2\rangle\langle0|$
without changing the reduced filter or feedback. For $\varepsilon>0$, $\Tr(\Pi_0\mathcal D_{C_{\mathrm{leak}}}(\Pi_0)) =-\varepsilon.$
Thus, \textbf{A2} fails and the joint target loses invariance.
Since the additional term $\varepsilon\Pi_0$ in $Q_C$ is diagonal,  \textbf{A8} remains valid.
For $T=2\tau$, Fig.~\ref{Fig:Leakage} compares
$\varepsilon=0.001$ and $0.01$ on $[0,200]$, using
$100$ trajectories and the same initial states and the controller as above.
Both cases exhibit initial target approach followed by departures,
with a larger late-time empirical mean distance for $\varepsilon=0.01$.
The simulations show that the feedback maintains high target
occupation under weak leakage, while increasing $\varepsilon$
leads to more frequent departures from the target and a larger
empirical mean distance. These results highlight the influence
of target invariance on the performance of reduced-filter-based
feedback stabilization.

%

%
%
%

\section{Conclusions and future work}
\label{Sec:conclusion}

This paper investigated robust feedback stabilization
of quantum systems subject to bounded measurable
time-varying Hamiltonian, dissipative, and measurement
uncertainties. Using an observation-driven reduced filter
with $d$ independent coordinates, we established
almost-sure exponential convergence of the physical
system and the reduced filter toward their joint target,
with an explicit sample exponent bound. The proposed
strategy requires neither knowledge of the initial state nor the exact time-dependent parameters,
relying instead on prescribed operator structures and
parameter bounds.
Our analysis identifies the operator-level mechanisms
underlying robust target accessibility. Joint
Hamiltonian--measurement reachability and a
noncancellation condition exclude persistence outside
the target, without requiring preservation of all QND
sectors. We also provided verifiable structural
criteria and established uniform finite-mean target
entrance through controlled accessibility, support
estimates, and a logarithmic exit argument. These
results extend the robustness analysis of reduced-filter
feedback to a broader class of nonautonomous quantum
dynamics.
Future work will investigate  extensions beyond QND measurements, and
feedback performance when perturbations destroy
exact target invariance.

\appendix
\renewcommand{\thetheorem}{\thesection.\arabic{theorem}}

\section{Proof of Theorem~\ref{Prop:Invariance_q}}\label{App:ProofThmInv_q}
First, suppose that $u_t$  is a real, bounded nonnegative progressively measurable process on the underlying filtration.
Under the coefficient bounds, the standard normalization argument for finite-dimensional SMEs
ensures the existence and the uniqueness of a global strong solution
$\rho(t)\in\mathcal S(\mathcal H)$; see
\cite[Section~3]{mirrahimi2007stabilizing}
and \cite[Chapter~5]{barchielli2009quantum}.
The argument applies to the present measurable
coefficients by taking $\sqrt{\theta_k}L_k$
as the observed operators and
$\sqrt{\gamma_k-\theta_k}L_k$, together with $C_r$,
as the unobserved operators.

Suppose first that $\hat q(0)\in\mathcal O_{d}$,
and let $\tau_e$ be the maximal lifetime of the
local estimator solution.  Since the coefficients are measurable and locally Lipschitz in the state,
the solution is pathwise unique on $[0,\tau_e)$.
Denote $S_t=\sum_{n=0}^d\hat q_n(t)$. Due to \textbf{A4}(i), by summing~\eqref{Eq:SDE}, we have 
\[
\textstyle d(1-S_t) =-2(1-S_t)\sum_k\sqrt{\hat\theta_k} \Lambda_k(\hat q) \bigl(dW_k+\mathfrak T_k(t,\rho,\hat q)\,dt\bigr).
\]
Since $S_0=1$, localization and uniqueness for
this homogeneous equation imply $S_t=1$
for all $t<\tau_e$ almost surely.

Choose $k_0\geq2$ sufficiently large so that $\hat{q}_n(0)>1/k_0$ for all $n\in\{0,\dots,d\}$. For each integer $k\geq k_0$, define $ \tau_k:=\inf\{t\in[0,\tau_e)|\min_n \hat{q}_n(t)\leq 1/k \}$ where $\tau_k<\tau_e$ whenever $\tau_e<\infty$.
Before $\tau_k\wedge\tau_e$, by the continuity and normalization, we have  $1/k<\hat q_n\leq1$. 
Define $$\bar{\mathcal O}_{d}^{k}:=\{q\in\bar{\mathcal O}_{d}|q_i\geq1/k\text{ for every }i\}.$$
Consider the function $V(\hat{q})=\textstyle\sum^d_{n=0}\log \hat{q}_{n}^{-1}\in\mathcal{C}^2(\mathcal{O}_{d},\mathbb{R}_+),$ whose infinitesimal generator is given by 
$$\mathscr{L}_tV(\hat q)=-u_t\mathcal{A}(\hat{q})+\mathcal{B}(t,\rho,\hat{q}),$$
where
\begin{align*}
&\mathcal{A}(\hat{q}):=\sum^d_{n=0}\frac{1}{\hat{q}_n}\sum^d_{j=0}\Gamma_{n,j}\hat{q}_j,\quad \mathcal{B}(t,\rho,\hat{q}):= 2\sum_{n,k}\sqrt{\hat{\theta}_k}\Phi^k_n(\hat{q})\big[\sqrt{\hat{\theta}_k}\Phi^k_n(\hat{q})-\mathfrak{T}_k(t,\rho,\hat{q})\big].
\end{align*}
Due to the compactness of $\mathcal X$, there exists  $c_1\geq0$ such that
$\mathcal B(t,\rho,\hat q)\leq c_1$ uniformly in time.
Rewrite
\[
\mathcal A(\hat q)=\operatorname{Tr}\Gamma
          +\sum_{n\ne j}\Gamma_{n,j}\frac{\hat q_j}{\hat q_n}
          \geq\operatorname{Tr}(\Gamma).
\]
With $0\leq u_t\leq\bar u$ and $V\geq(d+1)\log(d+1)$, it follows that
\[
\mathscr L_tV\leq c_1-\bar u\operatorname{Tr}\Gamma=:c
\]
for some finite constant $c\geq0$ independent of $k$. By applying It\^o's formula up to
$T\wedge\tau_k$, we have
\[
\mathbb E V(\hat q(T\wedge\tau_k))
\leq V(\hat q(0))+cT.
\]
On $\{\tau_k\leq T\}$, $\min_n\hat q_n(\tau_k)=k^{-1}$.
Since all coordinates are at most one, $V(\hat q(\tau_k))\geq\log k$, which implies
\[
\begin{aligned}
\log k\,\mathbb P(\tau_k\leq T)
&\leq
\mathbb E [V(\hat q(T\wedge\tau_k))]\leq V(\hat q(0))+c T.
\end{aligned}
\]
Then, we deduce $\mathbb P(\tau_k\leq T)\to 0$ as $k\to\infty$.
Since $\tau_k<\tau_e$ on
$\{\tau_e<\infty\}$, we have
\[
\mathbb P(\tau_e\leq T)
\leq\mathbb P(\tau_k\leq T)
\longrightarrow0 \quad\text{as }k\to\infty.
\]
Thus, $\tau_e=\infty$ almost surely. Furthermore, by continuity, any trajectory reaching the boundary of
$\mathcal O_{d}$ before time $T$ must reach the boundary of $\bar{\mathcal O}_{d}^{k}$ for every $k\geq k_0$.
Since $T>0$ is arbitrary, we conclude that $\hat q(t)\in\mathcal O_{d}$ for all $t\geq 0$ almost surely in the case where $u_t$ is a real, bounded, nonnegative progressively measurable process.

Now consider $u_t=u(\hat q_t)$.
Let $\mathcal E$ be a bounded open neighborhood of the compact
closed joint state space in its finite-dimensional real affine space.
Choose a smooth compactly supported cutoff $\vartheta$ equal to one
on $\mathcal E$, and use the bounded nonnegative Lipschitz extension
\(
U(q):=u\big(\operatorname{proj}_{\bar{\mathcal O}_{d}}q\big).
\)
Projection onto the closed simplex makes this extension globally Lipschitz.                                 
The cutoff equations are
\begin{align*}
\small
d\varrho&= \vartheta(\varrho,\hat{\mathfrak q})
 \Big[\big(\mathcal L^{U(\hat{\mathfrak q})}_\gamma(t,\varrho)
               +\mathcal P(t,\varrho)\big)dt\sum_k\sqrt{\theta_k(t)}\mathcal G_{L_k}(\varrho)dW_k\Big],\\
d\hat{\mathfrak q}_n&= \vartheta(\varrho,\hat{\mathfrak q})
 \Big[U(\hat{\mathfrak q})\sum_j\Gamma_{n,j}\hat{\mathfrak q}_jdt2\hat{\mathfrak q}_n\sum_k\sqrt{\hat\theta_k}
 \Phi_n^k(\hat{\mathfrak q})
       \big(dW_k+\mathfrak T_k(t,\varrho,\hat{\mathfrak q})dt\big)\Big].
\end{align*}
All arguments in these equations refer to the cutoff solution. Since the coefficients are globally Lipschitz in state and uniformly in time,  there is a unique global strong solution.
Set $u_t=U(\hat{\mathfrak q}_t)$ and solve the original uncut equations with this input and the same initial point and Wiener process. The first part of the proof gives a global solution in $\mathcal S(\mathcal H)\times\mathcal O_{d}$.
Both solutions coincide up to the first exit from $\mathcal E$ by
pathwise uniqueness, and the latter solution never leaves the compact
state space  $\mathcal X\subset \mathcal E$. Thus, the cutoff is inactive and the coupled equations have the unique global strong solution.

Finally, let $\hat q(0)\in\bar{\mathcal O}_{d}$ and set
$\hat q^{(n)}(0)=(1-n^{-1})\hat q(0)
+n^{-1}\mathbf1/(d+1)$.
Let $\widetilde Z^{(n)}$ and $\widetilde Z$ solve
the same cutoff equations with initial states
$(\rho(0),\hat q^{(n)}(0))$ and
$(\rho(0),\hat q(0))$, respectively, driven by
the same Wiener process.
Due to the global Lipschitz, for every finite $T>0$, we have 
\[
\textstyle \mathbb E\sup_{t\le T}\|\widetilde Z^{(n)}(t)- \widetilde Z(t)\|^2 \le C_T\|\hat q^{(n)}(0)-\hat q(0)\|^2\to0,
\]
as $n\to \infty$, where each $\widetilde Z^{(n)}$ remains in $\mathcal X$.
Since $\mathcal X$ is closed, the limit also remains
in $\mathcal X$ throughout $[0,T]$ almost surely.
Taking the countable intersection over integer
$T\geq1$ proves invariance for all $t\geq0$.
Hence, the cutoff is inactive for boundary initial
states as well. Global strong existence follows,
and pathwise uniqueness follows from the local
state-space Lipschitz bounds.
\hfill$\square$


\section{Support for measurable time coefficients}
\label{App:support}

By applying the classical support approach~\cite{millet1994simple}, together with the commuting QND diffusion fields under \textbf{A1}, and the Gaussian support and translation results in~\cite[Lemmas~5.1--5.2]{vandervaart2008rkhs}, we can obtain the following uniform support estimate.

\begin{lemma}
\label{lem:measurable_support}
Assume \textbf{A1}, \textbf{A4}(i), and
\textbf{A5}. Let $Z$ and $z_v$ solve
\eqref{Eq:SME}--\eqref{Eq:SDE} and
\eqref{Eq:ODE}--\eqref{Eq:ODE_F},
respectively, with the same initial condition
$z\in\mathcal X$.
For every $T,R,\varepsilon>0$, there exists
$p(T,R,\varepsilon)>0$ such that
\begin{equation*}
\mathbb P_z^{\mathfrak a} \left( \sup_{0\leq t\leq T} \|Z(t)-z_v(t)\|<\varepsilon \right) \geq p(T,R,\varepsilon)
\end{equation*}
uniformly for all $z\in\mathcal{X}$, $\mathfrak a\in\mathfrak U$, and $v\in L^2([0,T];\mathbb R^m)$ satisfying
$\int_0^T|v(t)|^2dt\leq R$.
\end{lemma}

\proof
Write $g_k(\rho)=\mathcal G_{L_k}(\rho)$ and
$f_{k,n}(\hat q)=2\hat q_n\Phi_n^k(\hat q)$. Define the joint
Gaussian driver $\Xi=(\Xi^\rho,\Xi^q)$ by
\[
   \Xi_k^\rho(t)=\int_0^t\sqrt{\theta_k(s)}\,dW_k(s),
   \qquad \Xi_k^q(t)=\sqrt{\hat\theta_k}\,W_k(t).
\]
The corresponding fields $(g_k,0)$ and $(0,f_k)$ are independent of
time. Under \textbf{A1}, they generate a commuting flow
$\varphi_\xi$ on $\mathcal X$. Indeed, for
$\xi=(\xi^\rho,\xi^q)\in\mathbb R^{2m}$ and
$E_\xi=\exp(\sum_k\xi_k^\rho L_k)$, its components are
\begin{align*}
 \varphi_\xi^\rho(\rho)
 &=\frac{E_\xi\rho E_\xi^*}{\Tr(E_\xi\rho E_\xi^*)},&
 \varphi_{\xi,n}^q(\hat q)
 &=\frac{\hat q_n\exp(2\sum_k\xi_k^q\Re\{l_{k,n}\})}
         {\sum_j\hat q_j\exp(2\sum_k\xi_k^q\Re\{l_{k,j}\})}.
\end{align*}
These formulas imply $\varphi_\xi\circ\varphi_\zeta=\varphi_{\xi+\zeta}$
and $\varphi_\xi^{-1}=\varphi_{-\xi}$, including on the boundary of
$\mathcal X$.

Let $b_{\mathfrak a}$ denote the joint It\^o drift. The Stratonovich
drift with respect to $\Xi$ is
\[
 b_{\mathfrak a}^{\circ}
  =b_{\mathfrak a}-\tfrac12\sum_k
       \big(\theta_k Dg_k\,g_k,\ \hat\theta_k Df_k\,f_k\big).
\]
The two coordinates driven by $W_k$ have covariation
$d\langle\Xi_k^\rho,\Xi_k^q\rangle_t
=\sqrt{\theta_k(t)\hat\theta_k}\,dt$. Its contribution vanishes
because the mixed directional derivatives of $(g_k,0)$ and $(0,f_k)$
are zero. Write $s_k(\rho)=\Tr((L_k+L_k^*)\rho)$. Differentiating
$g_k$ and $f_k$, and using normality of $L_k$ under \textbf{A1}, we obtain
\begin{align*}
 (b_{\mathfrak a}^{\circ})_\rho
 &=\tilde{\mathcal L}_{\gamma,\eta}^{u}+\mathcal P
       +\sum_k\theta_k s_k g_k,\quad (b_{\mathfrak a}^{\circ})_q
 =\mathfrak f^u+\Delta
       +\sum_k\sqrt{\hat\theta_k\theta_k}\,s_k f_k .
\end{align*}
Thus, adding the control $v_k$ in the two noise coordinates with their
respective amplitudes yields precisely the term
$V_k=v_k+\sqrt{\theta_k}\,s_k(\rho_v)$
in~\eqref{Eq:ODE} and~\eqref{Eq:ODE_F}.

For $y\in\mathcal X$, define the pullback of the corrected drift by
\[
 F_{\mathfrak a}(t,\xi,y)
   =D\varphi_{-\xi}(\varphi_\xi(y))\,
       b_{\mathfrak a}^{\circ}(t,\varphi_\xi(y)),
\]
where $D$ denotes the state derivative in the affine span of
$\mathcal X$. For $|\xi|\leq B$, the denominators in the flow formulas
are bounded away from zero uniformly on $\mathcal X$. The flow and
its state derivative are  Lipschitz in $(\xi,y)$ on this set.
The common coefficient bounds and the $C^1$ regularity of the fixed
feedback in \textbf{A5} give a uniform state Lipschitz bound for
$b_{\mathfrak a}^{\circ}$. Hence, $F_{\mathfrak a}$ is measurable in
$t$, bounded, and Lipschitz in $(\xi,y)$, with constants
independent of $\mathfrak a$ and of $t$ outside a null set.

Let $P_{\mathcal X}$ be the Euclidean metric projection onto the compact convex
set $\mathcal X$. For each continuous path $\xi$ starting at zero, let
$y^\xi(0)=z$ and solve
\[
 \dot y^\xi(t)=F_{\mathfrak a}(t,\xi(t),P_{\mathcal X} y^\xi(t)).
\]
Since $P_{\mathcal X}$ is nonexpansive, this equation has a unique global solution
on $[0,T]$. By Gronwall's inequality, we deuce
\[
 \|y^\xi-y^{\tilde\xi}\|_\infty
 \leq C_{T,B}\|\xi-\tilde\xi\|_\infty
 \quad\text{if}\quad
 \max\{\|\xi\|_\infty,\|\tilde\xi\|_\infty\}\leq B ,
\]
uniformly in $\mathfrak a$ and $z$. By Theorem~\ref{Prop:Invariance_q},
$Z$ takes values in $\mathcal X$. Therefore, the Stratonovich chain rule with
respect to the semimartingale $\Xi$ identifies
$y^\Xi(t)=\varphi_{-\Xi(t)}(Z(t))\in\mathcal X$ almost surely on
$[0,T]$.

The deterministic shift $h_v=(h_v^\rho,h_v^q)$ corresponding to $v$ is
\[
 h_{v,k}^\rho(t)=\int_0^t\sqrt{\theta_k(s)}\,v_k(s)\,ds,
 \qquad
 h_{v,k}^q(t)=\sqrt{\hat\theta_k}\int_0^t v_k(s)\,ds .
\]
Indeed, $h_v(t)=\mathbb E[\Xi(t)\int_0^T v(s)^\top dW(s)]$.
Therefore, by Gaussian covariance representation, we have 
$h_v\in\mathcal H_\Xi$ and
$\|h_v\|_{\mathcal H_\Xi}^2\leq\int_0^T|v(s)|^2\,ds\leq R$.
In particular, $h_v$ belongs to the support of the law of $\Xi$
by~\cite[Lemma~5.1]{vandervaart2008rkhs}.
By continuity, the set of drivers $\xi$ for which
$y^\xi([0,T])\subset\mathcal X$ is closed. It has full measure
under the law of $\Xi$, and contains $h_v$. Thus,
$P_{\mathcal X} y^{h_v}=y^{h_v}$. The ordinary chain rule now shows that
$\varphi_{h_v(t)}(y^{h_v}(t))$ solves~\eqref{Eq:ODE}
and~\eqref{Eq:ODE_F}. Its drift has a state Lipschitz bound of the
form $C(1+|v(t)|)$, integrable on $[0,T]$, so uniqueness identifies
this path with $z_v$.

The measurement bounds give $\|h_v\|_\infty\leq C_\theta\sqrt{TR}$.
Take $B=C_\theta\sqrt{TR}+1$. The preceding Gronwall bound and the
Lipschitz bound for $\varphi$ give a $\delta\in(0,1)$, depending
only on $T,R,\varepsilon$ and the fixed data and bounds, such that
$\|\Xi-h_v\|_\infty<\delta$ implies
$\|Z-z_v\|_\infty<\varepsilon$ almost surely. Applying the Gaussian
translation inequality~\cite[Lemma~5.2]{vandervaart2008rkhs} to the
symmetric open ball of radius $\delta$, we obtain
\begin{align}
 \mathbb P_z^{\mathfrak a}(\|Z-z_v\|_\infty<\varepsilon)
 &\geq \mathbb P(\|\Xi-h_v\|_\infty<\delta)\geq e^{-R/2}\mathbb P(\|\Xi\|_\infty<\delta).
 \label{eq:uniform_support_translation}
\end{align}

Finally, $\Xi(0)=0$ and
$\mathbb E|\Xi(t)-\Xi(s)|^4\leq C|t-s|^2$ uniformly in
$\mathfrak a$. Therefore, its laws are tight on the space of
continuous $\mathbb R^{2m}$-valued paths starting at zero, equipped
with the supremum norm. Their weak closure is compact, and every
law in the closure is centered Gaussian, as follows from its
finite-dimensional distributions. Each such law gives positive mass
to the open ball about zero of radius $\delta$
by~\cite[Lemma~5.1]{vandervaart2008rkhs}. By the Portmanteau theorem,
the mass of this open ball is lower semicontinuous in the law.
Thus, it has a strictly positive minimum $p_0$ on the compact
closure. Equation~\eqref{eq:uniform_support_translation} implies the
claim with $p(T,R,\varepsilon)=e^{-R/2}p_0$. The same bound applies
after any nonnegative time shift since $\mathfrak U$ is closed under
such shifts.\hfill$\square$

\bibliographystyle{IEEEtran}
\bibliography{IEEEabrv,ref_IEEEabrv}

\end{document}